\pdfoutput=1
\documentclass[twocolumn,showkeys,showpacs,preprintnumbers,prd,superscriptaddress,nofootinbib]{revtex4-1}
\usepackage{graphicx}
\usepackage{epsf}
\usepackage{bm}
\usepackage{amsmath}
\usepackage{amsfonts}
\usepackage{amssymb}
\usepackage{epstopdf}
\usepackage{natbib}
\usepackage{hyperref}
\usepackage{color}
\usepackage{verbatim}
\usepackage{multirow}
\usepackage{bm}
\usepackage{hyperref}
\usepackage{url}
\usepackage{orcidlink}
\usepackage{wrapfig}
\usepackage{nicefrac, xfrac}
\usepackage{enumitem} 
\usepackage{cleveref}
\usepackage[normalem]{ulem}

\definecolor{navyblue}{rgb}{0.0, 0.0, 0.5}
\definecolor{royalblue}{rgb}{0.25, 0.41, 0.88}
\definecolor{cadmiumgreen}{rgb}{0.0, 0.42, 0.24}
\definecolor{blue-violet}{rgb}{0.54, 0.17, 0.89}
\definecolor{darkviolet}{rgb}{0.58, 0.0, 0.83}
\definecolor{orange(colorwheel)}{rgb}{1.0, 0.5, 0.0}

\usepackage{hyperref}
\hypersetup{
    colorlinks=true, 
    linkcolor=royalblue, 
    citecolor=magenta}

\makeatletter\let\expandableinput\@@input\makeatother

\definecolor{WildStrawberry}{HTML}{EE2967}

\begin{document}

\title{Neutrino mass tension or suppressed growth rate of matter perturbations?}

\author{William Giar\`e \orcidlink{0000-0002-4012-9285}}
\email{w.giare@sheffield.ac.uk}
\affiliation{School of Mathematical and Physical Sciences, University of Sheffield, Hounsfield Road, Sheffield S3 7RH, United Kingdom}

\author{Olga Mena \orcidlink{0000-0001-5225-975X}}
\email{omena@ific.uv.es}
\affiliation{Instituto de F\'isica Corpuscular (IFIC), University of Valencia-CSIC, Parc Cient\'ific UV, c/ Catedr\'atico Jos\'e Beltr\'an 2, E-46980 Paterna, Spain}

\author{Enrico Specogna \orcidlink{0009-0005-6764-2555}}
\email{especogna1@sheffield.ac.uk}
\affiliation{School of Mathematical and Physical Sciences, University of Sheffield, Hounsfield Road, Sheffield S3 7RH, United Kingdom}

\author{Eleonora Di Valentino \orcidlink{0000-0001-8408-6961}}
\email{e.divalentino@sheffield.ac.uk}
\affiliation{School of Mathematical and Physical Sciences, University of Sheffield, Hounsfield Road, Sheffield S3 7RH, United Kingdom}

\begin{abstract}

Assuming a minimal $\Lambda$CDM cosmology with three massive neutrinos, the joint analysis of Planck cosmic microwave background data, DESI baryon acoustic oscillations, and distance moduli measurements of Type Ia supernovae from the Pantheon+ sample sets an upper bound on the total neutrino mass, $\sum m_\nu \lesssim 0.06$–$0.07$~eV, that lies barely above the lower limit from oscillation experiments. These constraints are mainly driven by mild differences in the inferred values of the matter density parameter across different probes that can be alleviated by introducing additional background-level degrees of freedom (e.g., by dynamical dark energy models). However, in this work we explore an alternative possibility. Since both $\Omega_\mathrm{m}$ and massive neutrinos critically influence the growth of cosmic structures, we test whether the neutrino mass tension may originate from the way matter clusters, rather than from a breakdown of the $\Lambda$CDM expansion history. To this end, we introduce the growth index $\gamma$, which characterizes the rate at which matter perturbations grow. Deviations from the standard $\Lambda$CDM value ($\gamma \simeq 0.55$) can capture a broad class of models, including non-minimal dark sector physics and modified gravity. We show that allowing $\gamma$ to vary significantly relaxes the neutrino mass bounds to $\sum m_\nu \lesssim 0.13$–$0.2$~eV, removing any tension with terrestrial constraints without altering the inferred value of $\Omega_\mathrm{m}$. However, this comes at the cost of departing from standard growth predictions: to have $\sum m_\nu \gtrsim 0.06$~eV one needs $\gamma > 0.55$, and we find a consistent preference for $\gamma > 0.55$ at the level of $\sim 2\sigma$. This preference increases to $\sim 2.5$–$3\sigma$ when a physically motivated prior $\sum m_\nu \ge 0.06$~eV from oscillation experiments is imposed.
\end{abstract}

\maketitle

\section{Introduction}

What began as Pauli's desperate remedy to preserve energy conservation in beta decay -- the postulation of light, electrically neutral, and almost undetectable particles, soon to be called neutrinos -- has since grown into one of the most remarkable success stories in modern physics. Despite Pauli's own skepticism about the testability of his hypothesis, not only have three distinct families of neutrinos been experimentally confirmed, but neutrino oscillation experiments have also precisely measured two independent mass-squared differences: $|\Delta m^2_{31}| \equiv |m_3^2 - m_1^2| \simeq 2.5 \times 10^{-3}\,\mathrm{eV}^2$ and $\Delta m^2_{21} \equiv m_2^2 - m_1^2 \simeq 7.5 \times 10^{-5}\,\mathrm{eV}^2$~\cite{deSalas:2020pgw,Esteban:2024eli,Capozzi:2025wyn}.

On the one hand, these measurements imply that at least two neutrino species have nonzero mass, providing concrete evidence for physics beyond the Standard Model, where neutrinos are originally treated as massless particles. Explaining neutrino masses -- either through the effective dimension-five Weinberg operator~\cite{Weinberg:1979sa}, which induces small Majorana masses by violating lepton number, or via more elaborate seesaw mechanisms that introduce heavy right-handed neutrinos~\cite{Minkowski:1977sc,Mohapatra:1979ia,Gell-Mann:1979vob,King:2025eqv} -- remains an open challenge that attracts significant research interest. 

On the other hand, since the sign of $\Delta m^2_{31}$ remains unknown, two possible mass orderings are still allowed: the \emph{normal ordering} (NO) $\Delta m^2_{31} > 0$ and the \emph{inverted ordering} (IO) $\Delta m^2_{31} < 0$. In both orderings, assuming the lightest neutrino mass is zero, neutrino oscillation experiments provide a lower limit on the total neutrino mass, $\sum m_\nu \equiv m_1 + m_2 + m_3$. This reads $\sum m_\nu > 0.06$~eV for the NO and $\sum m_\nu > 0.1$~eV for the IO~\cite{Esteban:2024eli}.

Given their intrinsic importance,\footnote{For example, neutrinoless double beta decay searches, which investigate the possible Majorana nature of neutrinos (namely, whether neutrinos are their own antiparticles), depend crucially on both the absolute neutrino mass scale and their hierarchical ordering~\cite{Deppisch:2012nb,DellOro:2016tmg}.} determining the neutrino mass ordering and precisely measuring the total neutrino mass $\sum m_\nu$ remain key objectives of neutrino physics. To this end, a number of ongoing and future neutrino oscillation experiments have been designed to resolve the mass ordering while simultaneously improving constraints on $\sum m_\nu$~\cite{T2K:2023smv, NOvA:2021nfi, JUNO:2015zny, Hyper-Kamiokande:2018ofw, DUNE:2020ypp}; see also Ref.~\cite{Gariazzo:2018pei}. Complementing these efforts, direct searches such as the KATRIN experiment provide competitive upper bounds on the total neutrino mass, currently $\sum m_\nu < 1.35$~eV~\cite{Katrin:2024tvg}. 

On the cosmological side, in the simplest scenario -- that is, within the standard $\Lambda$ Cold Dark Matter ($\Lambda$CDM) model and three degenerate neutrino states -- the DESI collaboration has recently reported that when DESI-2025 Baryon Acoustic Oscillation (BAO) measurements are combined with Planck-PR4 Cosmic Microwave Background (CMB) likelihoods, the cosmological upper limit on the total neutrino mass reads $\sum m_\nu \lesssim 0.064$~eV~\cite{DESI:2025ejh} at 95\% Confidence Level (CL). This is worryingly close to the lower limits set by oscillation experiments within the NO and in strong tension with the IO. Furthermore, when extending the analysis to other sources of cosmological and astrophysical data, the cosmological bound on the total neutrino mass can be pushed even below the minimum scale predicted by terrestrial oscillation experiments, i.e., around $\sum m_\nu \lesssim 0.05$~eV~\cite{Wang:2024hen}, leading to an overall tension between oscillation results and cosmological observations~\cite{Jiang:2024viw}.\footnote{Note that pre-2024 cosmological constraints on $\sum m_\nu$ ranged from the conservative 95\% CL limit $\sum m_\nu \lesssim 0.2$~eV quoted by the Planck Collaboration in Ref.~\cite{Planck:2018vyg} all the way up to the tightest bounds $\sum m_\nu \lesssim 0.09$~eV~\cite{Palanque-Delabrouille:2019iyz, DiValentino:2021hoh, Brieden:2022lsd} resulting from joint analyses of Planck PR3 CMB data and late-time expansion history probes in the form of distance measurements from BAO and Type Ia Supernovae (SNIa). While the most constraining results placed the IO under some tension, the exact preference for the NO over the IO (and more broadly the tightest limits themselves) varied significantly depending on the number of underlying assumptions and datasets involved in the different analyses. Therefore, before DESI BAO data, discussions surrounding preferences for specific orderings and tensions in neutrino cosmology were somewhat academic in nature.} Although part of this tension may reflect differences or inconsistencies across datasets, an even more striking feature emerges when the well-motivated physical prior $\sum m_\nu > 0$ (used in the DESI analyses) is relaxed, and the unphysical region $\sum m_\nu < 0$ is allowed. In this case, negative neutrino masses are found to be favored by the data, as reported by several independent groups~\cite{Loverde:2024nfi,Craig:2024tky, Naredo-Tuero:2024sgf, Green:2024xbb, Elbers:2024sha}.

It now appears well established that, within the minimal $\Lambda$CDM framework, the extremely small values inferred for the total neutrino mass largely originate from a mismatch in the preferred value of $\Omega_m$~\cite{Loverde:2024nfi,Colgain:2024mtg,DESI:2025ejh,Lynch:2025ine,Sailer:2025lxj,Jhaveri:2025neg}. In particular, BAO distance measurements from DESI favor a lower matter density, which lies in moderate ($2$-$3\sigma$) tension with values inferred from CMB observations. This tension often manifests as a preference for unnaturally small values of $\sum m_\nu$.\footnote{It is worth noting that, assuming a $\Lambda$CDM cosmology, this tension must necessarily originate from systematic effects responsible for the discrepancies among datasets. In addition to the possibility of undetected systematics in DESI BAO measurements~\cite{Wang:2024pui,Colgain:2024xqj,Naredo-Tuero:2024sgf,Sapone:2024ltl,Colgain:2025fct,Colgain:2025nzf,Giare:2025pzu,Efstathiou:2025tie}, another relevant factor is the prominent role played by CMB data~\cite{Giare:2024oil}. Even before the release of the DESI results, some of us had already shown that temperature and polarization measurements particularly at large scales can be affected by anomalies that influence constraints on parameters tightly correlated with $\sum m_{\nu}$, such as $\tau$ and $A_{\rm lens}$~\cite{Giare:2023ejv}. Following the DESI results, several independent analyses confirmed that relaxing the constraints from large-scale $E$-mode polarization and allowing $A_{\rm lens}$ to vary can significantly weaken the bounds quoted earlier~\cite{RoyChoudhury:2024wri,Gariazzo:2024sil,Sailer:2025lxj,Jhaveri:2025neg,RoyChoudhury:2025dhe,Capozzi:2025wyn}.} 

Interestingly, when considering extended cosmological scenarios -- most notably those allowing for dynamical dark energy -- the discrepancy is somewhat alleviated. As originally shown by the DESI collaboration~\cite{DESI:2024mwx,DESI:2025zgx,DESI:2025wyn} and subsequently corroborated by multiple re-analyses~~\cite{Cortes:2024lgw,Shlivko:2024llw,Luongo:2024fww,Yin:2024hba,Gialamas:2024lyw,Dinda:2024kjf,Najafi:2024qzm,Wang:2024dka,Ye:2024ywg,Tada:2024znt,Carloni:2024zpl,Chan-GyungPark:2024mlx,DESI:2024kob,Bhattacharya:2024hep,Ramadan:2024kmn,Notari:2024rti,Orchard:2024bve,Hernandez-Almada:2024ost,Pourojaghi:2024tmw,Giare:2024gpk,Reboucas:2024smm,Giare:2024ocw,Chan-GyungPark:2024brx,Menci:2024hop,Li:2024qus,Li:2024hrv,Notari:2024zmi,Gao:2024ily,Fikri:2024klc,Jiang:2024xnu,Zheng:2024qzi,Gomez-Valent:2024ejh,RoyChoudhury:2024wri,Lewis:2024cqj,Wolf:2025jlc,Shajib:2025tpd,Giare:2025pzu,Chaussidon:2025npr,Kessler:2025kju,Pang:2025lvh,RoyChoudhury:2025dhe,Scherer:2025esj,Specogna:2025guo,Teixeira:2025czm,Cheng:2025lod,Cheng:2025hug,Ozulker:2025ehg,Gialamas:2025pwv,Fazzari:2025lzd}, adopting a Chevallier Polarski Linder (CPL) parametrization~\cite{Chevallier:2000qy, Linder:2002et}, in which the dark energy equation of state varies linearly with the expansion history of the Universe, leads to a significantly improved consistency between DESI BAO and Planck CMB data. In this extended scenario, a preference for dynamical dark energy emerges at the $2.8$--$4.2\sigma$ level, depending on the specific dataset combination. At the same time, the neutrino mass bounds are relaxed up to $\sum m_\nu < 0.16$~eV, restoring agreement with oscillation experiments~\cite{DESI:2025zgx,Capozzi:2025wyn}.

Given the mounting evidence for dynamical dark energy and its resilience across diverse independent probes, the predominant -- though far from unanimous\footnote{With no intention to underplay the robustness and intrinsic interest of the hints for dynamical dark energy that emerged after DESI, we note that this interpretation is not without caveats. Both the DESI result and the preference for dynamical dark energy have been the subject of debate and several reinterpretations~\cite{Cortes:2024lgw,Colgain:2024mtg,Wang:2024pui,Colgain:2024xqj,Naredo-Tuero:2024sgf,Sapone:2024ltl,Efstathiou:2024xcq,Colgain:2025nzf,Giare:2024gpk,Giare:2024oil,Efstathiou:2025tie,Ye:2025ark}. Furthermore, while dynamical dark energy appears to be well supported by observations, it is unable to resolve or alleviate other major tensions in cosmology, such as the Hubble tension~\cite{Verde:2019ivm,DiValentino:2020zio,DiValentino:2021izs,Perivolaropoulos:2021jda,Schoneberg:2021qvd,Shah:2021onj,Abdalla:2022yfr,DiValentino:2022fjm,Kamionkowski:2022pkx,Giare:2023xoc,Hu:2023jqc,Verde:2023lmm,DiValentino:2024yew,Perivolaropoulos:2024yxv,CosmoVerse:2025txj}, leaving room for the exploration of more elaborate scenarios potentially capable of systematically accounting for all these problems simultaneously, see, e.g., Refs.~\cite{Allali:2024cji,Jiang:2024xnu,Giare:2024smz,Wang:2024dka,Wang:2024tjd,Poulin:2025nfb}.} -- narrative surrounding the neutrino mass problem is that the unphysically small cosmological upper limits on $\sum m_{\nu}$ may themselves reflect the inability of the minimal $\Lambda$CDM model to accurately describe the late-time expansion of the Universe. In other terms, within the minimal $\Lambda$CDM model, the late-time behavior of $H(z)$ is tightly governed by only two parameters: the present-day Hubble constant $H_0$ and the matter density parameter $\Omega_m$. Stringent constraints from DESI BAO data on the transverse comoving distance $D_M(z)$, the Hubble parameter $H(z)$, and their combination $D_V(z)$ (all expressed relative to the sound horizon at the drag epoch $r_d$) translate directly into stringent bounds on $\Omega_m$, driving the mismatch between cosmological upper limits and terrestrial lower limits on the total neutrino mass. Conversely, an evolving dark energy equation of state introduces additional freedom into $H(z)$, allowing more flexibility in fitting BAO measurements at different redshift while accommodating values of $\Omega_m$ that are in line with CMB-derived estimates. As a result, the differences in the inferred value of the matter density, combined with the enlarged parameter space, weaken the degeneracy between $\Omega_m$ and $\sum m_\nu$, relaxing the upper bounds on the neutrino mass.\footnote{We note that yet another key assumption underlying cosmological limits on the total neutrino mass is the standard neutrino physics. Many alternative models of new physics in the neutrino sector have been proposed and remain viable and actively explored alternatives, often leading to weaker cosmological bounds on $\sum m_{\nu}$. Without any claim of completeness, see, e.g., Refs.~\cite{Cuoco:2005qr, Farzan:2015pca, Dvali:2016uhn, Bellomo:2016xhl, Lattanzi:2017ubx, Lorenz:2018fzb, Kreisch:2019yzn, Oldengott:2019lke, Chacko:2019nej,RoyChoudhury:2018gay,RoyChoudhury:2018vnm,RoyChoudhury:2019hls, Chacko:2020hmh, Escudero:2020ped, Esteban:2021ozz, Esteban:2022rjk, FrancoAbellan:2021hdb, Dvali:2021uvk, Alvey:2021xmq, Alvey:2021sji, Lorenz:2021alz, DEramo:2022nvb, Escudero:2022gez, Sen:2023uga, Giare:2023qqn, Brax:2023rrf, Brax:2023tvn, Allali:2024anb,Loverde:2024nfi,Trojanowski:2025oro, Benso:2024qrg, Zu:2025lrk, Poudou:2025qcx,Das:2025asx}.} In this sense, one could argue that dynamical dark energy can be regarded as a solution that operates almost entirely at the level of \textit{background} dynamics, leaving minimal effects at the level of perturbations in the CMB spectra.

Although this interpretation has gained significant traction, in this work we wish to propose and explore a complementary perspective on the neutrino mass discrepancy by shifting the focus from the expansion history to the growth of cosmic structures. The key observation motivating this shift is that both $\Omega_m$ and $\sum m_{\nu}$ not only affect the background expansion but also critically influence the growth of cosmic structures, driving gravitational clustering of over-densities and suppressing structure formation at small scales, respectively. Consequently, a mismatch between cosmological and terrestrial bounds on the total neutrino mass may point to new physics impacting not only the background expansion but also the dynamics of the growth rate of matter perturbations over time, and structure formation more broadly -- either alongside or instead of modifications to the expansion history.

To test this possibility, we consider a non-standard growth of structure by introducing a free growth index parameter, $\gamma$~\cite{Linder:2005in, Linder:2007hg}. This index effectively captures the physics of structure growth across a wide range of alternative scenarios, from dark energy models to beyond-Einstein gravity. Crucially, $\gamma$ does not alter the background expansion. Instead, it directly impacts observables sensitive to matter perturbations, such as the CMB spectra and the matter power spectrum. This makes it possible to disentangle different models based on their perturbation-level effects, even when their expansion histories are identical. In this sense, it offers a clean framework to test whether new physics affecting the growth of structure can provide an alternative and complementary route to reconcile cosmological and terrestrial bounds on the neutrino mass. To this end, we perform a comprehensive analysis combining the latest DESI BAO data, Planck CMB likelihoods, and Type Ia supernova distance moduli measurements, explicitly allowing for departures from $\Lambda$CDM at the \textit{perturbation} level while keeping the background fixed to the baseline model.

The structure of the manuscript is as follows. We start in Sec.~\ref{sec:theory} by describing the linear growth factor and the growth index parameter, illustrating the degeneracy between the former and the neutrino mass. Section~\ref{sec:data} presents the different datasets employed in the analyses and the methodology followed to derive the results presented in Sec.~\ref{sec:results}. We draw our conclusions in Sec.~\ref{sec:conclusions}.

\section{Growth index and neutrinos}
\label{sec:theory}

\begin{figure*}[ht!]
   \includegraphics[width=0.85\textwidth]{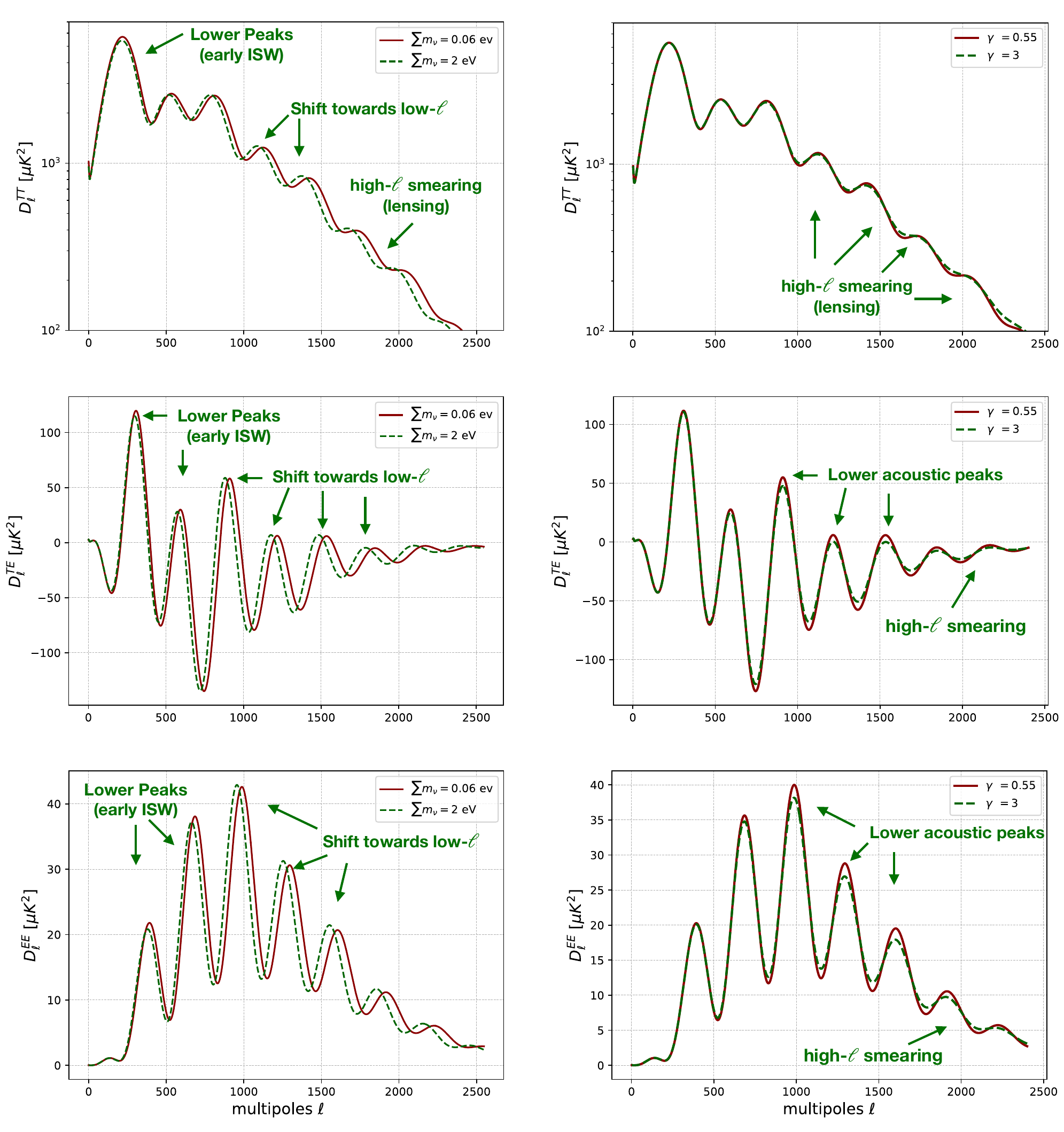}
   \caption{CMB temperature, polarization and cross-correlation power spectra for different values of the total neutrino mass  (left panels), and for different values of the growth index $\gamma$ (right panels), see text for details.}
   \label{fig:CMB}
\end{figure*}

\begin{figure}[ht!]
   \includegraphics[width=\columnwidth]{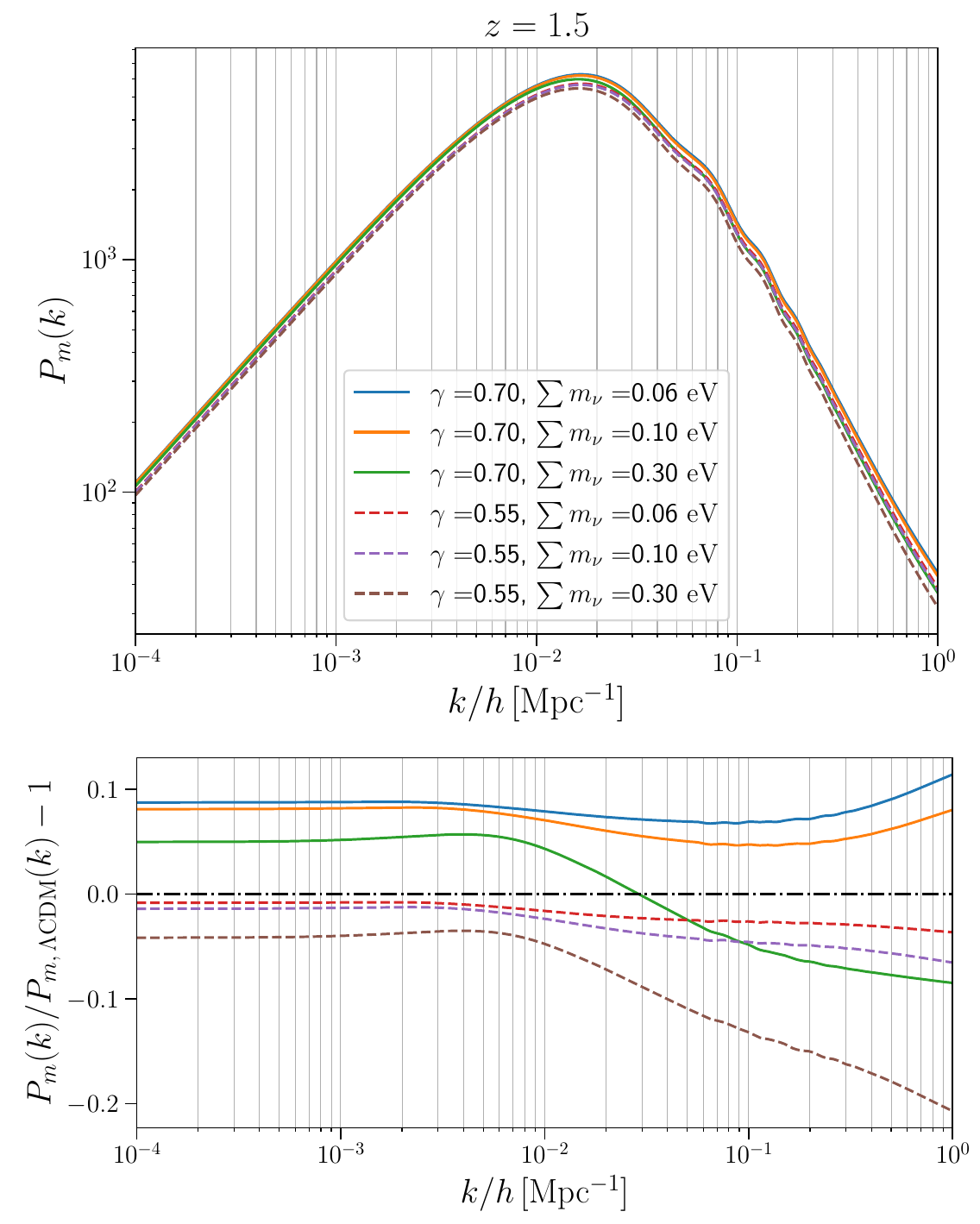} 
   \caption{Matter power spectrum at redshift $z = 1.5$ for a standard growth scenario ($\gamma = 0.55$, dashed lines) and a modified one with $\gamma = 0.7$ (solid lines). The different colors depict different values of the total neutrino mass, namely $\sum m_\nu = 0.06$, $0.1$, and $0.3$~eV. In the bottom panel, we show the ratios of the power spectra for the different parameter cases relative to the $\Lambda$CDM model.}
   \label{fig:matterpower}
\end{figure}

As emphasized in the introduction, our goal is to assess whether the unphysically tight bounds on the total neutrino mass, arising from the mild tension between the values of $\Omega_m$ preferred by geometric probes like BAO and those inferred from perturbation-sensitive observables such as the CMB in $\Lambda$CDM, may be driven, at least in part, by the assumption of standard structure growth, without necessarily indicating a failure of the standard background expansion, as in dynamical dark energy models. In this regard, many well-motivated extensions of the standard cosmological model predict a background evolution virtually identical to that of $\Lambda$CDM. Consequently, these models exhibit degeneracy in $\Omega_m(a)$ and cannot be distinguished based on background-level predictions alone. This happens, for instance, in those $\Lambda$CDM extensions where the effective equation of state of the accelerating component in the Hubble equation (whether this is dark energy or a modification of gravity) is free to vary slightly around $-1$  \cite{Linder:2016xer}. Similarly, in the DGP model \cite{Dvali:2000hr} one can reproduce the same expansion history predicted by fixing $w_0= -0.78$ and $w_a= 0.32$ in the CPL parametrization \cite{Linder:2005in}, making it hard to distinguish different models by means of background quantities only.
However, such models can still produce distinct signatures in structure formation. If the actual growth of structures deviates from the $\Lambda$CDM expectation, these deviations could partially mimic or offset the effects usually attributed to massive neutrinos. Hence, the evolution of matter perturbations offers a powerful probe of alternative scenarios, ranging from modified gravity theories to non-minimal dark sector physics.

The growth of density perturbations can be studied using the perturbed equations of motion derived from the underlying gravitational theory. In the sub-horizon, linear regime, the physical interpretation is straightforward: the growth is driven by a source term proportional to the amount of matter that can cluster, and suppressed by a friction term -- often referred to as Hubble drag -- associated with the expansion of the Universe~\cite{Ma:1995ey,Bernardeau:2001qr}.

A key observable in this regime is the growth rate of structure, defined as 
\begin{equation}
f = \frac{d\ln{\delta_\mathrm{m}}}{d\ln{a}}=\frac{\delta'_\mathrm{m}}{\delta_\mathrm{m}}a~,
\end{equation}
where $\delta_\mathrm{m}$ is the matter density contrast and $\prime$ denotes the derivative with respect to the scale factor. The growth rate captures how efficiently matter overdensities grow over time, from their primordial seeds to the observed large-scale structure.

A commonly used approach to capture deviations from standard structure formation history is to introduce the so-called growth index $\gamma$~\cite{Linder:2005in, Wang:1998gt}, defined as the exponent that yields a power-law solution to the growth rate equation in terms of the matter density parameter: 
\begin{equation}
\label{eq:growth_factor}
f(a) \simeq \Omega_m(a)^\gamma \simeq \left[\frac{\Omega_\mathrm{m}a^{-3}}{\Omega_\mathrm{de}(a)+\Omega_\mathrm{m}a^{-3}}\right]^\gamma~.
\end{equation}
Note that, in this expression, $\Omega_m(a)$ denotes the matter density fraction at scale factor $a$. 

Although $\gamma$ can, in general, acquire a time and/or scale dependence according to the underlying theory of gravity or dark energy, in many relevant cases it can be accurately approximated as a constant. For a dark energy fluid with constant equation of state $w$, one finds $\gamma \simeq 3(w - 1)/(6w - 5)$~\cite{Linder:2007hg}, which gives the canonical value $\gamma = 6/11 \simeq 0.55$ in the $\Lambda$CDM case ($w = -1$)~\cite{Linder:2005in}. In many alternative scenarios, however, $\gamma$ can differ substantially. Just to mention a few concrete examples: in DGP braneworld gravity, one finds $\gamma \approx 0.68$, with less than $2\%$ variation when treated as constant~\cite{Linder:2005in, Linder:2007hg}, while in $f(R)$ gravity models with mild scale dependence, $\gamma \approx 0.4$~\cite{Tsujikawa:2009ku}, though a significant time dependence may still arise, with present-day slopes as large as $\gamma' \simeq -0.2$~\cite{Gannouji:2008wt}.

Overall, the growth index plays a central role in linking theory to observations concerning the clustering of matter perturbations. Specifically, in linear perturbation theory, the evolution of matter inhomogeneities is governed by the linear growth function, defined as $D(a) \equiv \delta_{\mathrm{m}}(a) / \delta_{\mathrm{m}}(a_0)$, and satisfying $f(a) \equiv d \ln D(a) / d \ln a$. Assuming the parametrization in Eq.~\eqref{eq:growth_factor} with constant $\gamma$, we obtain~\cite{Nguyen:2023fip,Specogna:2023nkq}
\begin{equation}
\label{eq:growth_function}
D(\gamma,a) = \exp\left[ - \int_{a}^{1} \frac{\Omega_{\mathrm{m}}(a)^\gamma}{a} \, da \right],
\end{equation}
where the growth function is normalized such that $D(\gamma, a = 1) = 1$ for all values of $\gamma$. This introduces an important subtlety that will be crucial for interpreting the following results in the context of massive neutrino cosmologies. Namely, models predicting a larger growth index, i.e. $\gamma > 0.55$, imply a \textit{slower} growth rate at any given epoch. Indeed, since $\Omega_{\mathrm{m}}(a) < 1$ by construction, Eq.~\eqref{eq:growth_factor} implies $f(\gamma > 0.55, a) < f(\gamma = 0.55, a)$, corresponding to a lower formation rate. However, due to the normalization $D(a = 1) = 1$, a slower growth rate today implies that the linear growth function must have been \textit{larger} in the past to reach the same amplitude at present. In other words, we find that $D(\gamma > 0.55, a) > D(\gamma = 0.55, a)$ for $a < 1$. As a result, models with larger $\gamma$ predict an enhanced growth function in the past, despite a systematically reduced growth rate throughout cosmic history.

Keeping this result in mind, when it comes to massive neutrinos, many important observations are in order to clarify their impact on structure formation and their connection with the growth index $\gamma$. 

First and foremost, once neutrinos become non-relativistic particles, they contribute to the total non-relativistic matter density -- a contribution that depends ultimately on the value of the total neutrino mass. It is therefore unsurprising that different values of $\sum m_{\nu}$ would inherently affect the theoretical value of the growth index $\gamma$, even within the standard $\Lambda$CDM framework. An analytical expression for the growth rate $f$ in the presence of massive neutrinos was derived in Ref.~\cite{Kiakotou:2007pz}. However, we have numerically confirmed that, even for total neutrino masses well above the current observational limits, the theoretical variations in $\gamma$ remain negligible compared to the uncertainties in its inferred value. Consequently, having established that the theoretical impact of treating neutrinos as massive particles on $\gamma$ is effectively negligible for the purpose of this study, we will use $\gamma = 0.55$ as our reference value in the following analysis to test deviations from the $\Lambda$CDM model.

A second, more subtle but important point -- partially noted in the introduction -- is that both massive neutrinos and deviations from the canonical growth index value impact cosmological observables related to perturbations (such as the CMB angular power spectra and the matter power spectrum) in distinct yet qualitatively similar ways.

Massive neutrinos affect the angular power spectra of CMB temperature and polarization anisotropies through several mechanisms: by contributing to the late-time non-relativistic energy density, they alter the angular scale relation at the last scattering surface and modify the late Integrated Sachs-Wolfe (ISW) effect. Moreover, the exact value of $\sum m_\nu$ influences the neutrino transition to the non-relativistic regime by changing their pressure-to-density ratio, inducing metric fluctuations visible in the early ISW effect. Most significantly, massive neutrinos suppress the growth of structure, reducing the amplitude of the matter power spectrum. This suppression also affects the weak lensing of the CMB, damping power on small angular scales.\footnote{For constraints on the neutrino mass based on CMB lensing or damping tail effects, see Refs.~\cite{Giare:2023aix,DiValentino:2023fei}. } The overall impact on the CMB power spectra includes a reduction in the amplitude of the first acoustic peak (mainly due to the ISW effect) and also a shifting of higher acoustic peaks toward lower multipoles. These effects are illustrated in the left panel of Fig.~\ref{fig:CMB}, which compares two reference total neutrino mass values: $\sum m_{\nu} = 0.06$~eV (the minimum mass consistent with oscillation experiments and commonly assumed in $\Lambda$CDM) and an unphysically large mass of $\sum m_{\nu} = 2$~eV, chosen to clearly highlight the effects described above. 

Similarly, non-canonical values of the growth index $\gamma$ directly modify the rate at which matter perturbations grow over time, affecting the amplitude and evolution of matter density fluctuations, which in turn influence secondary anisotropies in the CMB through gravitational lensing. Roughly speaking, since the lensing potential is highly sensitive to the distribution and amplitude of matter fluctuations along the line of sight, a growth index different from the canonical $\gamma \approx 0.55$ will alter the clustering strength of matter, changing the gravitational potential wells that CMB photons traverse on their journey toward us. This modification of the gravitational potentials affects the lensing of the CMB by smoothing the acoustic peaks in the temperature and polarization power spectra. This smoothing occurs because lensing deflects the paths of CMB photons, mixing power between different multipoles and reducing the sharpness of the acoustic peaks. As shown in the right panel of Fig.~\ref{fig:CMB}, increasingly large values of $\gamma>0.55$ produce more pronounced smoothing at small angular scales (high multipoles) -- exactly where the temperature anisotropy spectrum becomes approximately linearly dependent on the lensing power spectrum~\cite{Lewis:2006fu}. This effect can be easily compensated by increasing the neutrino mass, since neutrinos, which are hot dark matter particles, reduce CMB lensing as they suppress clustering at small scales~\cite{Kaplinghat:2003bh,Lesgourgues:2005yv}, leading to a less pronounced smoothing at low multipoles. Consequently, a large, positive degeneracy is expected among these two parameters, as we will precisely see in the results of our numerical analyses when both $\gamma$ and $\sum m_\nu$ are free parameters of the model.

Another way to understand the similar effects of $\gamma$ and $\sum m_{\nu}$ is by examining their impact on the matter power spectrum. This key complementary observable (which is ultimately linked to the CMB power spectra, especially at high multipoles) provides a direct way to quantify how deviations in $\gamma$ and the total neutrino mass impact the amplitude and shape of matter fluctuations. Fig.~\ref{fig:matterpower} depicts the matter power spectrum at $z = 1.5$ for the standard structure formation scenario with $\gamma = 0.55$ and a modified one with $\gamma = 0.7$. Three different possible values for $\sum m_\nu$ have been considered, namely the smallest total neutrino mass consistent with particle physics experiments ($\sum m_\nu = 0.06$~eV) and two other possible cases, $\sum m_\nu = 0.1$~eV and $\sum m_\nu = 0.3$~eV. As expected, larger neutrino masses suppress the matter power spectrum. Neutrinos are hot dark matter relics with very large velocity dispersions, suppressing the growth of structure at scales smaller than their free-streaming scale when they turn non-relativistic. The amount of suppression depends on the precise value of the total neutrino mass.

On the other hand, in cosmologies with a non-standard growth rate, the linear matter power spectrum can be expressed as~\cite{Nguyen:2023fip,Specogna:2023nkq}:
\begin{equation}
\label{eq:matter_power}
P_{\mathrm{m}}(\gamma,k,a)= P_{\mathrm{m}}^{\Lambda\mathrm{CDM}}(k,a=1)\, D^2(\gamma,a),
\end{equation}
where $P_{\mathrm{m}}^{\Lambda\mathrm{CDM}}(k,a=1)$ is the fiducial linear matter power spectrum evaluated today, while $D(\gamma,a)$ is the growth function defined in Eq.~\eqref{eq:growth_function}. As previously pointed out, a larger value of the growth index $\gamma$ suppresses the growth rate $f$, while enhancing the growth function $D(\gamma,a)$ at earlier times. According to Eq.~\eqref{eq:matter_power}, this translates into an enhanced matter power spectrum in the past (as seen in Fig.~\ref{fig:matterpower} at $z=1.5$). As a result, a positive degeneracy between $\sum m_\nu$ and $\gamma$ is expected: the suppression of structure formation due to larger neutrino masses can be partially compensated by a larger value of $\gamma$, which boosts the amplitude of $D(\gamma,a)$ and hence of the matter power spectrum. This is the case illustrated in Fig.~\ref{fig:matterpower}, where the matter power spectrum obtained for $\sum m_\nu = 0.1$~eV and $\gamma = 0.55$, is nearly indistinguishable from the case of $\sum m_\nu = 0.3$~eV and $\gamma = 0.7$ , at least for $k/h>0.1\rm\, Mpc^{-1}$. It is therefore expected that modified growth scenarios with $\gamma > 0.55$ can easily accommodate larger neutrino masses, and this very interesting possibility should be carefully analyzed with the most recent cosmological observations, as we shall present in the following sections.

\section{Methodology and datasets}
\label{sec:data}

The considerations outlined in Sec.~\ref{sec:theory} motivate us to analyze extensions of the $\Lambda$CDM parameter space -- hereafter referred to as $\lambda$ -- by including $\gamma$ and $\sum m_{\nu}$ as additional free parameters.\footnote{Note that $\lambda$ includes the six standard model parameters, namely: the matter densities of baryons ($\Omega_{\rm b} h^2$) and dark matter ($\Omega_{\rm c} h^2$), the optical depth to reionization ($\tau$), the amplitude ($A_s$) and spectral index ($n_s$) of the primordial scalar fluctuations, and the angular size of the sound horizon at last scattering ($\theta_{\rm MC}$).} In particular, we consider the following extended parameter space combinations:
\begin{itemize}
    \item $\lambda \cup \{\sum m_{\nu}\}$,
    \item $\lambda \cup \{\gamma\}$,
    \item $\lambda \cup \{\sum m_{\nu},\, \gamma\}$.
\end{itemize}

The cosmology calculations have been performed with a patched version of the Boltzmann solver \texttt{CAMB}~\cite{Lewis:1999bs, Howlett:2012mh}, specifically adapted for the $\gamma$ parameterization of modified gravity: \texttt{CAMB\_GammaPrime\_Growth}~\cite{Nguyen:2023fip}. We stress that, while $\gamma$ can, in general, be time-dependent, in this paper we take it to be constant at all redshifts.

To efficiently explore the extended parameter spaces, we perform Markov Chain Monte Carlo (MCMC) analyses using the \texttt{Cobaya} sampler~\cite{Torrado:2020dgo}, varying the cosmological parameters within the flat priors listed in Tab.~\ref{tab:priors}. Note that, as explained in more detail in the next section, for the total neutrino mass we consider two relevant cases: one in which we impose a lower prior limit $\sum m_{\nu} > 0.06$ eV, based on oscillation experiment results assuming the NO, and another in which this lower limit is relaxed to $\sum m_{\nu} > 0$ eV. We ensure that all produced chains satisfy the Gelman Rubin convergence criterion, requiring a convergence threshold of $R < 0.01$. These chains are then analyzed using the \texttt{Getdist} package~\cite{Lewis:2019xzd} to derive constraints on the parameter space and generate the plots presented in Sec.~\ref{sec:results}.

The dataset employed in our analysis includes two different combinations of likelihood releases for the CMB spectra measured by the Planck satellite. In particular:
\begin{itemize}
    \item \textbf{Plik}: This includes the Planck-PR3 likelihood \texttt{plik}~\cite{Planck:2019nip} for the temperature (TT), polarization (EE), and temperature polarization cross-correlation (TE) spectra at $\ell > 30$, the \texttt{simall} likelihood~\cite{Planck:2018vyg} for E-mode polarization measurements at $\ell < 30$, and the \texttt{Commander} likelihood~\cite{Planck:2018yye} for temperature anisotropies at $\ell < 30$. The \texttt{plik} lensing likelihood~\cite{Planck:2018lbu} is also included in this combination.
    \item \textbf{Camspec}: This includes the \texttt{Camspec}~\cite{Rosenberg:2022sdy} likelihood for the TT, TE, and EE spectra at $\ell > 30$, based on the Planck-PR4 \texttt{NPIPE} CMB maps~\cite{Planck:2020olo}, always in combination with the \texttt{simall} likelihood~\cite{Planck:2018vyg} for E-mode polarization and the \texttt{Commander} likelihood~\cite{Planck:2018yye} for temperature anisotropies at $\ell < 30$.  In this combination, for CMB lensing, we use the Planck-PR4 \texttt{NPIPE} lensing likelihood~\cite{Carron:2022eyg,Planck:2020olo}.
\end{itemize}

Note that we consider these two independent combinations of likelihoods for very important reasons: over the years, several mild anomalies have been identified in the Planck-PR3 data, many of which can alter the constraints on the parameters of interest in this study. The most significant example is the excess smoothing of acoustic peaks observed in the Planck-PR3 spectra, which appears to be reduced in the PR4 likelihoods based on the \texttt{NPIPE} maps.\footnote{This effect can be accurately quantified through the phenomenological parameter $A_{\rm lens}$~\cite{Calabrese:2008rt}, which captures the amplitude of the lensing power spectrum inferred from the smoothing of acoustic peaks in the temperature and polarization spectra. Analyses based on the Plik PR3 spectra suggest that $A_{\rm lens}$ deviates from the expected baseline value ($A_{\rm lens} = 1$) at the level of about $2.8\sigma$~\cite{Planck:2018vyg,DiValentino:2015bja,Renzi:2017cbg,Domenech:2020qay}. In contrast, when using the temperature and polarization PR4 spectra from \texttt{Camspec}, the preference for $A_{\rm lens} > 1$ is reduced to below $1.7\sigma$~\cite{Rosenberg:2022sdy}, even if this comes at the cost of a moderate internal disagreement between the value of the angular size of the sound horizon, $\theta$, inferred from temperature and polarization data, as well as an overall worse global fit to the different spectra.} Several groups have highlighted that these differences in excess smoothing between the PR3 and PR4 high-multipole likelihoods can significantly impact constraints on the total neutrino mass~\cite{DiValentino:2013mt,Capozzi:2017ipn,DiValentino:2021imh,Capozzi:2021fjo,Jiang:2024viw,Allali:2024aiv,Naredo-Tuero:2024sgf,Capozzi:2025wyn}. Similarly, as first shown by some of us in Ref.~\cite{Specogna:2024euz} and subsequently confirmed by the DESI collaboration itself~\cite{Ishak:2024jhs,DESI:2024hhd}, the overall differences in smoothing between the Planck-PR3 and Planck-PR4 spectra also have important consequences for constraints on the growth index $\gamma$. Depending on which high-multipole CMB likelihood is used, CMB constraints on $\gamma$ can vary significantly, resulting in either better or worse agreement with the baseline $\Lambda$CDM prediction~\cite{Specogna:2024euz}. Given the significant impact of choosing either the Planck-PR3 or Planck-PR4 likelihoods on the constraints of both parameters of interest, it is not only important, but in fact essential, to consider both combinations in order to draw reliable conclusions.

\begin{table}[t]
    \centering
    \renewcommand{\arraystretch}{1.2}
    \resizebox{0.766 \columnwidth}{!}{%
    \begin{tabular}{l @{\hspace{2cm}} c}
        \hline\hline
        Parameter & Prior \\
        \hline
        $\Omega_{\rm b} h^2$       & $[0.005,\ 0.1]$ \\
        $\Omega_{\rm c} h^2$       & $[0.001,\ 0.99]$ \\
        $\tau$                     & $[0.01,\ 0.8]$ \\
        $n_s$                      & $[0.8,\ 1.2]$ \\
        $\log(10^{10} A_s)$        & $[1.61,\ 3.91]$ \\
        $100\,\theta_{\rm MC}$     & $[0.5,\ 10]$ \\
        $\sum m_\nu$ [eV]          & $[0,\ 5]$ \\
        $\sum m_\nu$ [eV] (NO)     & $[0.06,\ 5]$\\
        $\gamma$                   & $[0,\ 1]$ \\
    \hline\hline
    \end{tabular}
    }
\caption{Flat priors adopted for the parameters presented in Sec.~\ref{sec:data}. The label “NO" denotes an additional prior restriction on $\sum m_\nu$, accounting for the normal neutrino ordering constraints from particle physics experiments~\cite{Esteban:2024eli}.}
\label{tab:priors}
\end{table}

In addition to the CMB data outlined above, we consider geometrical probes of the late Universe in the form of precise constraints on the transverse comoving distance, the Hubble rate, and their combination (all relative to the sound horizon at the drag epoch), derived from BAO data and distance modulus measurements from SNIa. In particular, we consider:
\begin{itemize}
    \item \textbf{DESI}: This includes the isotropic and anisotropic BAO measurements from the second data release by the DESI collaboration, based on observations of over 14 million extragalactic objects such as galaxies and quasars~\cite{DESI:2025qqy}, as well as Lyman-$\alpha$ tracers~\cite{DESI:2025zpo}. These measurements are summarized in Table IV of Ref.~\cite{DESI:2025zgx}.
    \item \textbf{PP}: These are the distance moduli obtained from observations of 1550 SNIa up to $z = 2.26$ in the Pantheon+ sample, without the inclusion of Cepheid host distances~\cite{Brout:2022vxf}.
\end{itemize}

In closing, we would like to spend a few words on the choice of the SNIa sample adopted in this analysis. At present, three independent compilations are available -- Pantheon+~\cite{Brout:2022vxf}, DESy5~\cite{DES:2024tys,DES:2024upw,DES:2024hip}, and Union3~\cite{Rubin:2023ovl} -- each constructed from different observational datasets and calibration strategies. Including all of them would considerably inflate the number of combinations under consideration, making the analysis unnecessarily cumbersome and less easy to interpret. For this reason, we restrict our analysis to Pantheon+, which is typically regarded as the more conservative catalogue, yielding constraints more consistent with Planck $\Lambda$CDM results. In contrast, DESy5 tends to favor deviations, especially in dynamical dark energy models. However, the nature of this shift remains debated. For instance, Ref.~\cite{Efstathiou:2024xcq} links the tension with Planck $\Lambda$CDM cosmology to a $\sim$0.04 mag calibration offset in low-redshift DESy5 SNIa. In response to these findings, the DES collaboration, in Ref.~\cite{DES:2025tir}, argued that such an offset arises from improved modeling of intrinsic scatter and host galaxy properties, as well as differences in the selection functions used across different samples. While we do not take a stance on this issue, these considerations motivate our use of Pantheon+, as it offers a conservative, well-tested compilation, limits residual systematics, and avoids unnecessary complexity. Last but not least, Pantheon+ yields the tightest neutrino mass constraints across both $\Lambda$CDM and dynamical dark energy scenarios (see, e.g., Ref.~\cite{DESI:2025zgx} or Tab.~8 of Ref.~\cite{Capozzi:2025wyn}).\footnote{For example, combining CMB and DESI with Pantheon+ gives $\sum m_\nu < 0.117$ eV in the dynamical dark energy model, while DESy5 leads to $\sum m_\nu < 0.129$ eV~\cite{DESI:2025zgx}.}

\begin{table*}[tpb!]
\renewcommand{\arraystretch}{1.5}
\resizebox{1.\textwidth}{!}{
\begin{tabular}{ l  c  c  c  c  c  c }
\hline
  \textbf{Parameter} & \textbf{Plik+DESI+PP} & \textbf{Plik+DESI+PP (NO)} & \textbf{Camspec+DESI+PP} & \textbf{Camspec+DESI+PP (NO)}\\
\hline
$\Omega_b h^2$ &  $ 0.02260\pm 0.00014 $&$ 0.02264\pm 0.00014 $&$ 0.02236\pm 0.00013 $&$ 0.02240\pm 0.00013$
\\
$\Omega_c h^2$ &  $ 0.1171^{+0.0012}_{-0.00086} $&$ 0.1165^{+0.0010}_{-0.00079} $&$ 0.11740^{+0.00091}_{-0.00075} $&$ 0.11682^{+0.00081}_{-0.00073}$
\\
$100 \, \theta_{MC} $ &  $ 1.04123\pm 0.00029 $&$ 1.04129\pm 0.00029 $&$ 1.04101\pm 0.00024 $&$ 1.04106\pm 0.00023$
\\
$\tau$ &  $ 0.0500^{+0.0085}_{-0.0074} $&$ 0.0502^{+0.0084}_{-0.0074} $&$ 0.0490^{+0.0086}_{-0.0074} $&$ 0.0494^{+0.0085}_{-0.0074}$
\\
${\rm{ln}}(10^{10} A_s)$ &  $ 3.027^{+0.018}_{-0.016} $&$ 3.026^{+0.018}_{-0.016} $&$ 3.024^{+0.018}_{-0.016} $&$ 3.023^{+0.018}_{-0.016}$
\\
$n_s$ &  $ 0.9726\pm 0.0038 $&$ 0.9739\pm 0.0037 $&$ 0.9696\pm 0.0037 $&$ 0.9711\pm 0.0036$
\\
$\gamma$ &  $ 0.707\pm 0.075$&$  0.742\pm 0.069 $&$ 0.660\pm 0.063 $&$ 0.696\pm 0.060$
\\
$\sum m_{\nu}$ [eV] &  $ < 0.188 $&$ < 0.208 $&$ < 0.134 $&$< 0.164$
\\
\hline
$\Omega_m$ &  $ 0.2993\pm 0.0040 $&$ 0.3000\pm 0.0040 $&$ 0.3004\pm 0.0039 $&$ 0.3016\pm 0.0038$
\\
$H_0$ [km/s/Mpc] &  $ 68.52\pm 0.35 $&$ 68.42\pm 0.34 $&$ 68.36\pm 0.32 $&$ 68.21\pm 0.31$
\\
$S_8$ &  $ 0.791^{+0.021}_{-0.014} $&$ 0.780^{+0.017}_{-0.013} $&$ 0.798^{+0.016}_{-0.012} $&$ 0.787^{+0.014}_{-0.011}$
\\
\hline
\end{tabular}
}
\caption{Mean values and $68\%$~CL errors on the most relevant cosmological parameters, including the growth index $\gamma$, together with the $95\%$~CL upper limits on the total neutrino mass $\sum m_{\nu}$ arising from different combinations of cosmological datasets.}
\label{tab:results_main}
\end{table*}

\section{Results}
\label{sec:results}

We present our main findings in Tab.~\ref{tab:results_main}, which compares the results obtained by simultaneously varying the neutrino mass and the growth index $\gamma$ using different CMB likelihoods in combination with DESI BAO and Pantheon+ SNIa measurements. For reference, results obtained by varying each of these parameters individually while keeping the other fixed are reported in Tab.~\ref{tab:results_appendix}, and discussed in \hyperref[Appendix-A]{Appendix A}. As already noted in the previous section, we consider two distinct prior choices for $\sum m_\nu$, as listed in Tab.~\ref{tab:priors}. Specifically, the label “NO” refers to the case where we impose the lower bound on $\sum m_\nu$ from particle physics experiments, i.e., enforcing a prior $\sum m_\nu > 0.06$~eV, consistent with current neutrino oscillation measurements~\cite{Esteban:2024eli}.

First and foremost, we confirm that the constraints on the neutrino mass are significantly relaxed compared to the minimal $\Lambda$CDM+$\sum m_{\nu}$ case with the growth index fixed to $\gamma = 0.55$. When both $\sum m_\nu$ and $\gamma$ are allowed to vary simultaneously, the most constraining bound, $\sum m_\nu < 0.134$~eV at 95\%~CL, is obtained using the Camspec likelihood in combination with DESI and PP data. This limit remains substantially larger -- by more than a factor of 2 -- than the corresponding bound obtained for the same dataset combination when fixing $\gamma = 0.55$, which yields $\sum m_\nu < 0.0643$~eV at 95\%~CL, remarkably close to the lower bound expected from terrestrial measurements.

Secondly, we confirm a non-negligible dependence on the specific CMB likelihood adopted in the analysis. As shown in Tab.~\ref{tab:results_main}, when varying both parameters, the bound on the neutrino mass changes from the aforementioned $\sum m_\nu < 0.134$~eV to $\sum m_\nu < 0.188$~eV upon replacing the Planck-PR4 Camspec and lensing likelihoods with the older Plik PR3 counterparts, further relaxing the upper limit on the total neutrino mass.

\begin{figure*}[ht!]
   \includegraphics[width=0.9\textwidth]{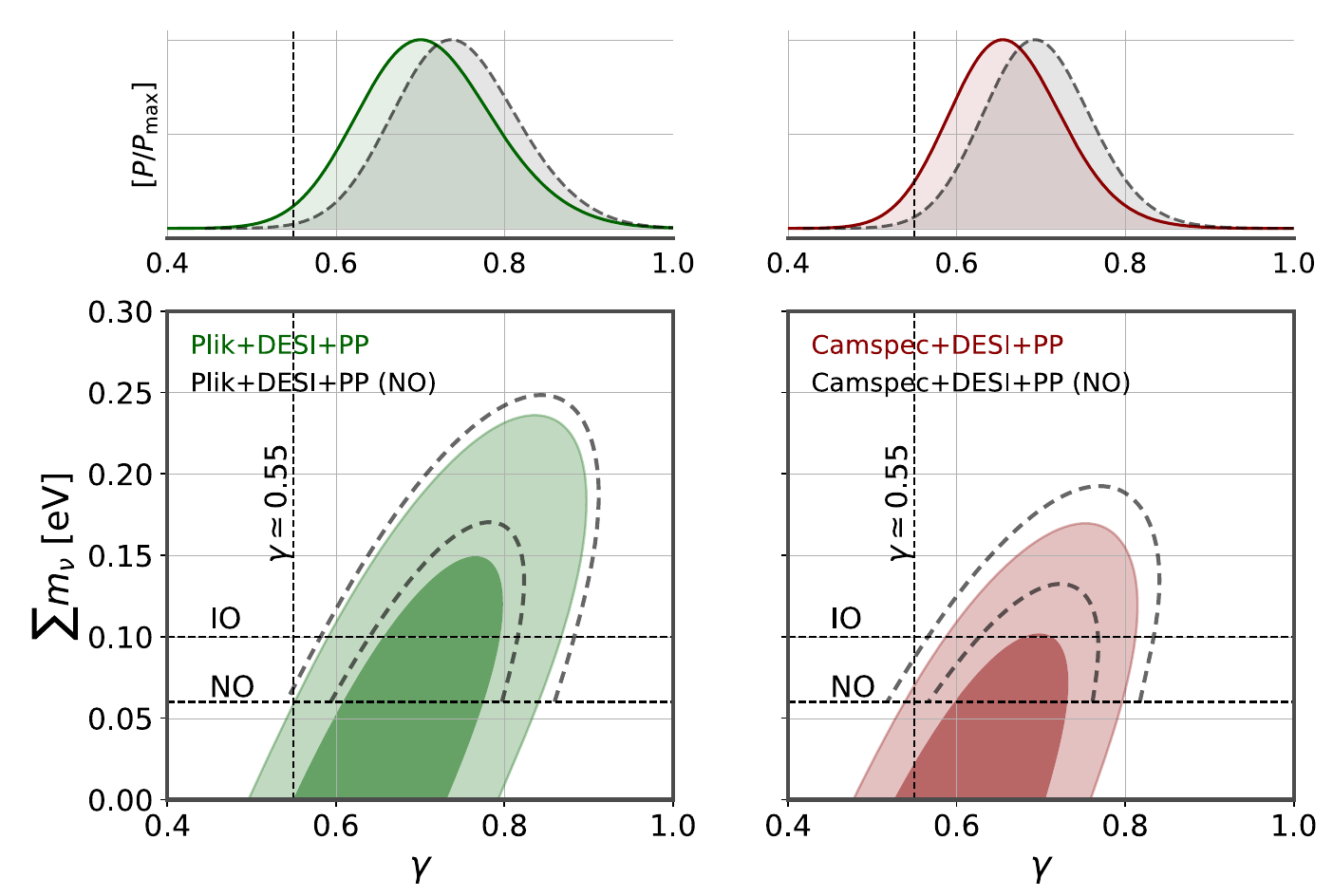}
     \caption{One-dimensional marginalized posterior distributions and two-dimensional 68\% and 95\%~CL contours in the ($\gamma$, $\sum m_\nu$) plane for different Planck CMB likelihoods combined with DESI and PP data. The horizontal black dashed line labeled “NO” marks the lower bound on the total neutrino mass set by oscillation experiments ($\sum m_\nu > 0.06$ eV). Values below this line are therefore in tension with oscillation data. Instead, the horizontal black dashed line labeled “IO” indicates the lowest value of the total neutrino mass compatible with the inverted ordering ($\sum m_\nu > 0.1$ eV). If the total neutrino mass lies between the NO and IO lines, only the normal ordering is allowed, whereas for values above the IO line both orderings remain viable. The vertical black dashed line corresponds to the standard prediction for the growth index ($\gamma \simeq 0.55$).}
    \label{fig:gmnu}
\end{figure*}

Either way, however, the limits on $\sum m_\nu$ remain largely consistent with the lower bounds from oscillation experiments, both within the NO and the IO, thereby alleviating the emerging tension observed in $\Lambda$CDM. Therefore we can relatively safely apply the NO prior in the analysis and enforce $\sum m_\nu > 0.06$~eV. Doing so, the neutrino mass bound from Camspec+DESI+PP becomes further relaxed to $\sum m_\nu < 0.164$~eV, implying a shift of about $0.03$~eV towards larger mass values. Similarly, when enforcing the NO prior, the bound obtained from Plik+DESI+PP becomes $\sum m_\nu < 0.208$~eV, corresponding to a shift of approximately $0.02$~eV. For both cases, the size and magnitude of this shift can be seen in Fig.~\ref{fig:gmnu}, which shows the 2D 68\% and 95\% probability contours in the ($\gamma$, $\sum m_\nu$) plane. Specifically, in the left panel, the solid green contours correspond to the results obtained from Plik+DESI+PP, while the empty dashed grey contours illustrate how these results shift when a NO prior is imposed. Similarly, the solid red contours in the right panel show the constraints from Camspec+DESI+PP, and the grey contours indicate how they change when enforcing $\sum m_\nu > 0.06$~eV. Clearly, the impact of the prior is more pronounced in the Camspec-based results.

Most importantly, as clearly seen from the two-dimensional allowed regions in Fig.~\ref{fig:gmnu}, there is a strong positive correlation between $\gamma$ and $\sum m_\nu$: larger values of the total neutrino mass imply a suppression of the growth of structure in the Universe, which can be compensated by a higher growth index. Therefore, when it comes to the growth index $\gamma$, several interesting observations are in order. Firstly, using the Camspec likelihood in combination with DESI and SNIa observations (without assuming a NO prior on the neutrino mass), we obtain $\gamma = 0.660 \pm 0.063$ at 68\%~CL, indicating a $\sim 1.8\sigma$ shift from the standard growth rate model. This shift increases to $\sim 2.1\sigma$ when replacing Camspec with Plik, which yields $\gamma = 0.707 \pm 0.075$ at 68\%~CL. Due to the strong positive correlation between $\gamma$ and $\sum m_\nu$, this tendency towards a larger growth index becomes significantly more pronounced when enforcing a NO prior on $\sum m_\nu$. For Camspec+DESI+PP we obtain $\gamma = 0.696 \pm 0.060$, while for Plik+DESI+PP we get $\gamma = 0.742 \pm 0.069$ -- i.e., $\sim 2.4\sigma$ and $\sim 2.8\sigma$ away from the expected value of $\gamma = 0.55$, respectively. This further shift towards larger values of $\gamma$ induced by the NO prior can be clearly seen in the 1D posterior distributions shown in the top histograms of Fig.~\ref{fig:gmnu}. In both the left and right panels, the dashed grey curves (corresponding to the analysis with the NO prior) are visibly shifted towards higher values of $\gamma$ compared to the distributions obtained when this prior is not applied.

The situation becomes even more intriguing when examining the 2D contours in Fig.~\ref{fig:gmnu} for the different cases. As already noted, allowing for a non-standard growth of structure can reconcile the current tension in the neutrino sector: even in the most constraining scenario, the cosmological bounds on neutrino masses remain fully compatible with laboratory results, showing no conflict between cosmological and terrestrial constraints. However, this agreement comes at the cost of departing from the standard structure growth predicted in $\Lambda$CDM. As illustrated in the figure, the point defined by the intersection of $\gamma = 0.55$ (i.e., the baseline prediction for the growth index) and $\sum m_\nu = 0.06$~eV (i.e., the minimal mass allowed by oscillation experiments) lies literally at the boundary of the 95\% confidence region for Plik+DESI+PP (i.e., the combination allowing the most freedom in the neutrino mass), and very close to the boundary for Camspec+DESI+PP. This means that any neutrino mass value larger than this lower limit would necessarily require a larger $\gamma$, thus moving away from $\Lambda$CDM predictions for structure formation. Conversely, if we wish to remain consistent with $\gamma = 0.55$ (i.e., following the vertical black dashed line in the figure), the only viable region lies below $\sum m_\nu = 0.06$~eV, thereby reintroducing the tension with oscillation experiments. Similarly, as far as the mass ordering is concerned, while the 95\%~CL upper bounds on $\sum m_\nu$ remain consistent with both NO and IO, crossing the lower limit set by oscillation experiments within the IO necessarily implies a departure from standard structure formation growth within this framework. Indeed, it is crucial to note that the point defined by $\gamma = 0.55$ and $\sum m_\nu = 0.1$~eV (i.e., the minimal mass allowed within the IO) always lies well outside the 95\% probability contours and consistency with the IO (i.e., restricting attention to the region of the plane above the corresponding horizontal line in Fig.~\ref{fig:gmnu}) necessarily implies values of $\gamma$ significantly larger than those predicted by $\Lambda$CDM. 

\begin{figure*}[ht!]
   \includegraphics[width=0.7\textwidth]{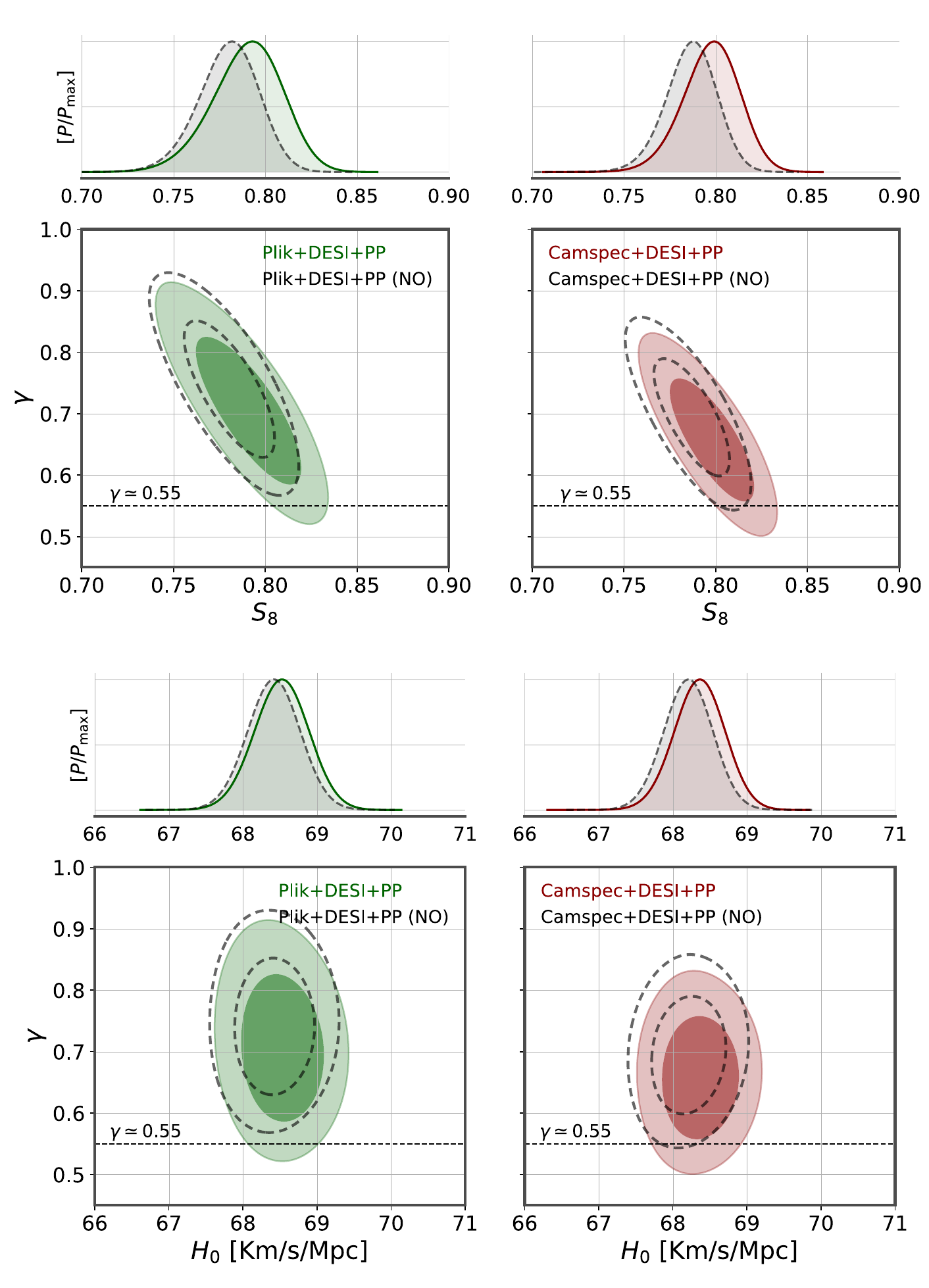}
   \caption{One-dimensional marginalized posterior distributions and two-dimensional 68\% and 95\% CL probability contours for $\gamma$ and $S_8$ (top panels) and $\gamma$ and $H_0$ (bottom panels) for the different Planck CMB likelihoods combined with DESI and PP. The horizontal black dashed line corresponds to the standard prediction for the growth index, $\gamma \simeq 0.55$.}
   \label{fig:tensions}
\end{figure*}

We now turn to a broader exploration of the implications of allowing departures from the standard structure growth history, focusing on additional parameters of interest for other outstanding cosmological tensions. We begin with the $S_8 \equiv \sigma_8 (\Omega_\mathrm{m}/0.3)^{0.5}$ parameter that quantifies the amplitude of matter fluctuations on intermediate scales and which has played a central role in the longstanding discussions surrounding potential discrepancies between CMB-derived results and weak lensing survey estimates~\cite{DiValentino:2020vvd,DiValentino:2018gcu,Nunes:2021ipq,DES:2021bvc,DES:2021vln,KiDS:2020suj,Asgari:2019fkq,Joudaki:2019pmv,DAmico:2019fhj,Kilo-DegreeSurvey:2023gfr,Troster:2019ean,Heymans:2020gsg,Dalal:2023olq,Chen:2024vvk,ACT:2024okh,DES:2024oud,Harnois-Deraps:2024ucb,Dvornik:2022xap,DES:2021wwk,Wright:2025xka}. As shown in Fig.~\ref{fig:tensions}, we observe a strong anti-correlation between $\gamma$ and $S_8$. Larger values of the growth index tend to favour both larger values of the total neutrino mass (see Fig.~\ref{fig:gmnu}) and smaller values of $S_8$. Quantitatively, for Plik+DESI+PP, we find $S_8 = 0.791^{+0.021}_{-0.014}$, which reduces to $S_8 = 0.780^{+0.017}_{-0.013}$ when assuming a NO prior. Similarly, for the Camspec likelihood, we obtain $S_8 = 0.798^{+0.016}_{-0.012}$ without prior assumptions, shifting to $S_8 = 0.787^{+0.014}_{-0.011}$ under the NO prior. In both cases, as seen in Fig.~\ref{fig:tensions}, the NO prior leads to a noticeable drift of the posteriors toward lower $S_8$ values and away from the canonical GR limit $\gamma = 0.55$, echoing the behaviour already discussed in previous sections. This interplay between $\gamma$, $\sum m_{\nu}$ and $S_8$ can be particularly relevant in light of ongoing discussions about the persistence of the weak lensing discrepancy, generalizing the discussion presented in Refs.~\cite{Nguyen:2023fip,Specogna:2023nkq} to massive neutrino cosmologies. Despite recent claims suggesting this tension may have been alleviated or even resolved~\cite{Stolzner:2025htz,Wright:2025xka}, our results indicate that the inclusion of a flexible growth history introduces parameter shifts in a direction potentially capable of reconciling lensing data with CMB-derived inferences, should the discrepancy persist. If we aim to be more quantitative in assessing the extent to which varying $\gamma$ alleviates the $S_8$ tension, we should first note that weak-lensing surveys report somewhat different determinations of $S_8$. To mention a few concrete examples, KiDS-1000 in combination with 2dFLenS and BOSS yields $S_8=0.766^{+0.020}_{-0.014}$~\cite{Heymans:2020gsg}, the DES-Y3 3$\times$2pt analysis gives $S_8=0.776\pm0.017$~\cite{DES:2021wwk}, while the more recent KiDS-Legacy release finds $S_8=0.815^{+0.016}_{-0.021}$~\cite{Wright:2025xka}, which, as mentioned above, shows no discrepancy with the $\Lambda$CDM Planck-only based inferences of $S_8=0.834\pm0.016$~\cite{Planck:2018vyg}.\footnote{For a more comprehensive review of $S_8$ measurements from different weak-lensing surveys, see Ref.~\cite{CosmoVerse:2025txj}.} A direct quantification of the tension is therefore non-trivial, not only because of these differing estimates but also because $S_8$ is itself a model-dependent parameter. As such, a fully consistent comparison would require reanalyzing the weak-lensing data within the same theoretical framework adopted here, which lies well beyond the scope of this manuscript. Nevertheless, even accounting for these caveats, one can clearly see that the results presented in \hyperref[Appendix-A]{Appendix~A}, Tab.~\ref{tab:results_appendix}, obtained for fixed $\gamma=0.55$, yield values scattered around $S_8\simeq0.816$ that, considering the uncertainties, remain in a $\sim2.1$–$2.3\sigma$ tension with KiDS-1000+2dFLenS+BOSS and DES-Y3, while being in good agreement with KiDS-Legacy. By contrast, when allowing the growth index $\gamma$ to vary freely, the inferred values shift to $S_8\simeq0.79$–$0.80$, consistently lying within $\lesssim1\sigma$ of all weak-lensing determinations. This heuristic argument shows that a flexible growth history can naturally drive the inferred clustering amplitude toward substantially improved consistency with lensing observations. More broadly, this drift underscores the relevance of neutrino physics assumptions in any interpretation of the weak lensing tension, and the model dependence of derived parameters such as $S_8$. Taken together, these findings suggest that $S_8$ remains sensitive to beyond-$\Lambda$CDM physics, particularly when degrees of freedom such as modified growth or non-canonical neutrino properties are considered.

In contrast to the behaviour observed for $S_8$, we do not find any significant correlation between $\gamma$ and $H_0$ in our analysis. The values we obtain remain remarkably stable across datasets and assumptions. For Plik+DESI+PP we find $H_0 = 68.52 \pm 0.35$~Km/s/Mpc, shifting only slightly to $H_0 = 68.42 \pm 0.34$~Km/s/Mpc when a NO prior is imposed. Similarly, for Camspec+DESI+PP we obtain $H_0 = 68.36 \pm 0.32$~Km/s/Mpc (without prior assumptions), decreasing mildly to $H_0 = 68.21 \pm 0.31$~Km/s/Mpc under the NO prior. These minor shifts can be traced back to the known negative correlation between $H_0$ and $\sum m_{\nu}$, and are not driven by the growth index itself. As such, while an alternative structure growth model with a free $\gamma$ can relax the neutrino mass bounds and might remain of interest in relation to the weak lensing discrepancy (should this issue persist rather than fade definitively) it does not provide a mechanism for addressing the Hubble tension. The path toward a coherent explanation capable of simultaneously accounting for all current anomalies remains, at present, largely unfulfilled.

Last but not least, it is important to consider the matter density parameter $\Omega_\mathrm{m}$, as it plays a key role in the interpretation of our results and lies at the heart of the broader discussion on the neutrino mass tension. As seen by comparing the values given in Tab.~\ref{tab:results_main} (as well as in Tab.~\ref{tab:results_appendix} in \hyperref[Appendix-A]{Appendix A}) across all the parameter spaces explored and for all combinations of datasets, we find values of $\Omega_\mathrm{m}$ that are consistent within one standard deviation. Specifically, when varying only the total neutrino mass, we find $\Omega_\mathrm{m} = 0.3009 \pm 0.0037$ for Plik+DESI+PP and $\Omega_\mathrm{m} = 0.3018 \pm 0.0036$ for Camspec+DESI+PP. When varying only the growth index $\gamma$, these values remain essentially unchanged: $\Omega_\mathrm{m} = 0.2990 \pm 0.0038$ (Plik+DESI+PP) and $\Omega_\mathrm{m} = 0.3006 \pm 0.0037$ (Camspec+DESI+PP).\footnote{We note that the tendency toward larger values of the growth index $\gamma$ is also present when fixing the total neutrino mass in the cosmological model, and that it persists regardless of the CMB likelihood employed; see the discussion in \hyperref[Appendix-A]{Appendix A}.} Finally, when both $\gamma$ and $\sum m_\nu$ are varied simultaneously, we obtain consistent values such as $\Omega_\mathrm{m} = 0.2993 \pm 0.0040$ (Plik+DESI+PP) and $\Omega_\mathrm{m} = 0.3004 \pm 0.0039$ (Camspec+DESI+PP). Under the NO prior, these latter two values shift only mildly to $\Omega_\mathrm{m} = 0.3000 \pm 0.0040$ and $\Omega_\mathrm{m} = 0.3016 \pm 0.0038$, respectively. These results illustrate a crucial point: fixing or varying $\gamma$ alongside the total neutrino mass does not significantly alter the inferred value of $\Omega_\mathrm{m}$. However, this consistency hides an important distinction. Fixing $\gamma = 0.55$ leads to neutrino mass bounds that are very close to lower limits from particle physics experiments, to the extent of raising a very concerning tension. Conversely, allowing $\gamma$ to vary relaxes this problem \textit{without} changing the value of $\Omega_\mathrm{m}$. This finding directly challenges the often-stated claim that the neutrino mass tension originates from constraints on $\Omega_\mathrm{m}$ inferred from different probes under the assumption of a $\Lambda$CDM background. We show that this is true if one also assumes a standard $\Lambda$CDM structure formation history. Once this assumption is relaxed, the tension in $\sum m_\nu$ can be alleviated without altering the background evolution. More broadly, our results demonstrate that it is possible to relax constraints on the total neutrino mass by modifying the structure growth history, without changing the inferred values of parameters describing the background dynamics and without invoking additional degrees of freedom at the background level, such as scenarios involving dynamical dark energy. This offers a complementary path to addressing the neutrino mass problem, shifting the attention to the important imprints of neutrinos on perturbation-level observables rather than on the background expansion itself. Exploring this interplay between perturbation-level and background-level modifications is crucial, as changes to the expansion history that help relax neutrino mass bounds often come at the cost of lowering $H_0$ further, thereby exacerbating the Hubble tension. If the goal is to retain enough freedom to simultaneously address multiple anomalies within a coherent framework, it becomes essential to introduce the necessary degrees of freedom and explore both aspects. In this sense, alternative structure formation histories may offer a viable and physically well-motivated possibility.

\section{Summary}
\label{sec:conclusions}

Assuming the minimal $\Lambda$CDM framework with three degenerate massive neutrinos, the DESI collaboration recently reported an upper bound of $\sum m_\nu \lesssim 0.064$~eV at 95\%~CL when combining DESI-2025 BAO data with Planck-PR4 CMB observations. This bound is uncomfortably close to the lower limit set by oscillation experiments in the case of normal ordering and firmly excludes the inverted ordering, thus raising serious concerns. In particular, it has been argued that these tight limits mainly stem from a mild but persistent mismatch in the preferred matter density: BAO measurements from DESI favour lower values of $\Omega_\mathrm{m}$ than those inferred from the CMB, resulting in an indirect pressure toward smaller values of $\sum m_\nu$.

An appealing way to alleviate this issue is to consider extensions beyond the standard cosmological framework. Models featuring dynamical dark energy, which modify the late-time expansion history of the Universe, can significantly relax the neutrino mass bounds and reduce the tension. Therefore, the mainstream (but far from universally accepted) interpretation is that the unnaturally low bounds on $\sum m_\nu$ may themselves point to the failure of $\Lambda$CDM to accurately describe the expansion history at late times. However, matter density and massive neutrinos do not only affect the background evolution, but also play crucial roles in the growth of cosmic structure: matter density drives the gravitational clustering of overdensities, while massive neutrinos suppress the formation of structure at small scales due to their large thermal velocities. Therefore, new physics affecting the dynamics of primordial perturbations and structure formation can offer a complementary possibility to alleviate the neutrino mass tension.

In this paper, we explore this possibility by considering a flexible, phenomenological approach to structure growth by allowing the growth rate $f(a)$ to depart from its $\Lambda$CDM prediction. Specifically, we parametrize it using the growth index $\gamma$ through $f(a) \simeq \Omega_\mathrm{m}(a)^\gamma$ and treat $\gamma$ as a free parameter. A larger value of $\gamma$ suppresses the growth rate, implying a larger growth function in the past and modifying the way matter perturbations cluster in the (early) Universe. This, in turn, affects both the high-multipole CMB lensing signal and the shape of the matter power spectrum. Such effects can be compensated by a larger neutrino mass. In this regard, it is important to emphasize the role of CMB lensing in driving the neutrino mass bounds, particularly when the growth index is fixed to its standard value. The lensing potential encodes the integrated matter distribution along the line of sight and is especially sensitive to the amplitude of clustering on intermediate and small scales. Since massive neutrinos suppress structure growth, their effect is clearly imprinted in the lensing signal at high multipoles. In the standard scenario with $\gamma = 0.55$, any deviation from the measured lensing power is necessarily attributed to a lower neutrino mass, thereby tightening the constraint on $\sum m_\nu$. However, when $\gamma$ is allowed to vary, structure suppression can be partially absorbed by a larger growth index, weakening the need for extremely small neutrino masses. This illustrates how perturbation-level flexibility, especially in structure growth, can mitigate tensions that appear under stricter assumptions.

Our main results can be summarized as follows:

\begin{itemize}

\item By jointly varying $\sum m_\nu$ and the growth index $\gamma$, we find that the cosmological bounds on the total neutrino mass are substantially relaxed compared to the standard case. Across all combinations of Planck-CMB, DESI-BAO, and Pantheon+ SNIa data, we find that allowing $\gamma$ to vary broadens the allowed range for $\sum m_\nu$ from $\sim 0.06$–$0.07$~eV (assuming $\gamma = 0.55$) up to $\sim 0.13$–$0.2$~eV. These values are consistent with terrestrial constraints and demonstrate that new physics at the perturbation level offers a viable, complementary solution to the neutrino mass tension.

\item This agreement, however, comes at the cost of deviating from the standard structure growth predictions of $\Lambda$CDM. As shown in Fig.~\ref{fig:gmnu}, the point defined by $\gamma = 0.55$ (i.e., the baseline value of $\gamma$) and $\sum m_\nu = 0.06$~eV (i.e., the lowest total neutrino mass allowed by particle physics experiments) lies on the boundary of the 95\% confidence region. This means that any neutrino mass value larger than this lower limit would necessarily require a larger $\gamma$, thus moving away from $\Lambda$CDM predictions for structure formation. Conversely, if we wish to remain consistent with $\gamma = 0.55$, the only viable region lies below $\sum m_\nu = 0.06$~eV, reintroducing the tension with oscillation experiments.
    
\item Regarding the neutrino mass ordering, we find that the inverted ordering remains allowed within modified structure growth models. However, accommodating mass values consistent with the inverted ordering requires values of $\gamma$ significantly larger than the $\Lambda$CDM prediction.
    
\item Overall, we observe a consistent trend favouring larger-than-standard values of $\gamma$. This preference persists whether one varies only $\gamma$ or both $\gamma$ and $\sum m_\nu$ simultaneously. In the latter case, the preference reaches $1.8\sigma$ for Plik+DESI+PP and $2\sigma$ for Camspec+DESI+PP. Imposing a prior consistent with normal ordering ($\sum m_\nu > 0.06$~eV) strengthens the preference further, up to $2.4\sigma$ and $2.8\sigma$, respectively.

\item Importantly, we find that allowing $\gamma$ to vary relaxes the neutrino mass bounds without altering the inferred value of $\Omega_\mathrm{m}$. This is clearly seen by comparing the values reported in Tab.~\ref{tab:results_main} and Tab.~\ref{tab:results_appendix}. While previous studies attributed the tight neutrino mass bounds to the ability of $\Lambda$CDM to accurately describe the late-time background dynamics, we show that this interpretation only holds under the assumption of standard $\Lambda$CDM structure growth. Once this assumption is relaxed, the tension can be alleviated without modifying the $\Lambda$CDM background evolution.
    
\item Last but not least, Fig.~\ref{fig:tensions} reveals a strong anti-correlation between $\gamma$ and $S_8$: larger growth indices lead to both larger allowed neutrino masses and smaller values of $S_8$. This interplay may be relevant for the ongoing debate about the weak lensing anomaly. While some recent analyses suggest that the $S_8$ tension may be subsiding, our results indicate that a flexible growth history can still shift parameters in a direction that could better reconcile lensing and CMB data, if the tension persists. On the other hand, we find no significant correlation between $\gamma$ and $H_0$, implying that growth index freedom does not offer a mechanism for addressing the Hubble tension.

\end{itemize}

All in all, in the absence of a definitive cosmological detection of neutrino mass, our analysis provides a novel and complementary angle on the tension between cosmology and particle physics. It highlights the critical role of perturbations in constraining new physics and underscores the importance of testing both the background and structure formation histories when interpreting cosmological data. As future observations improve and our modelling of structure formation becomes more refined, non-canonical scenarios (e.g., modified theories of gravity or a non-minimal dark sector) may well emerge as essential components of a more complete cosmological paradigm that accurately accounts for massive neutrinos.

\begin{acknowledgments}
\noindent W.G. is supported by the Lancaster Sheffield Consortium for Fundamental Physics under STFC grant: ST/X000621/1. 
O.M. acknowledges the financial support from the MCIU with funding from the European Union NextGenerationEU (PRTR-C17.I01) and Generalitat Valenciana (ASFAE/2022/020). 
E.D.V. is supported by a Royal Society Dorothy Hodgkin Research Fellowship. This work has been supported also by the Spanish MCIN/AEI/10.13039/501100011033 grants PID2020-113644GB-I00, ID2023-148162NB-C21 and
PID2023-148162NB-C22, and by the European ITN project HIDDeN (H2020-MSCA-ITN-2019/860881-HIDDeN) and SE project ASYMMETRY (HORIZON-MSCA-2021-SE-01/101086085-ASYMMETRY) and well as by the Generalitat Valenciana grants PROMETEO/2019/083, PROMETEO/2021/083, and CIPROM/2022/69.  OM acknowledges the financial support from the MCIU with funding from the European Union NextGenerationEU (PRTR-C17.I01) and Generalitat Valenciana (ASFAE/2022/020) We acknowledge the IT Services at The University of Sheffield for the provision of services for High Performance Computing. This article is based upon work from the COST Action CA21136 - ``Addressing observational tensions in cosmology with systematics and fundamental physics (CosmoVerse)'', supported by COST - ``European Cooperation in Science and Technology''.
\end{acknowledgments}

\bibliography{main}

@article{deSalas:2020pgw,
    author = "de Salas, P. F. and Forero, D. V. and Gariazzo, S. and Mart{\'\i}nez-Mirav{\'e}, P. and Mena, O. and Ternes, C. A. and T{\'o}rtola, M. and Valle, J. W. F.",
    title = "{2020 global reassessment of the neutrino oscillation picture}",
    eprint = "2006.11237",
    archivePrefix = "arXiv",
    primaryClass = "hep-ph",
    doi = "10.1007/JHEP02(2021)071",
    journal = "JHEP",
    volume = "02",
    pages = "071",
    year = "2021"
}

@article{Esteban:2024eli,
    author = "Esteban, Ivan and Gonzalez-Garcia, M. C. and Maltoni, Michele and Martinez-Soler, Ivan and Pinheiro, Jo{\~a}o Paulo and Schwetz, Thomas",
    title = "{NuFit-6.0: updated global analysis of three-flavor neutrino oscillations}",
    eprint = "2410.05380",
    archivePrefix = "arXiv",
    primaryClass = "hep-ph",
    reportNumber = "IFT-UAM/CSIC-24-140, YITP-SB-2024-24, IPPP/24/64, IPPP/24/64, IFT-UAM/CSIC-24-140, YITP-SB-2024-24",
    doi = "10.1007/JHEP12(2024)216",
    journal = "JHEP",
    volume = "12",
    pages = "216",
    year = "2024"
}

@article{Specogna:2023nkq,
    author = "Specogna, Enrico and Di Valentino, Eleonora and Levi Said, Jackson and Nguyen, Nhat-Minh",
    title = "{Exploring the growth index {\ensuremath{\gamma}}L: Insights from different CMB dataset combinations and approaches}",
    eprint = "2305.16865",
    archivePrefix = "arXiv",
    primaryClass = "astro-ph.CO",
    doi = "10.1103/PhysRevD.109.043528",
    journal = "Phys. Rev. D",
    volume = "109",
    number = "4",
    pages = "043528",
    year = "2024"
}

@article{Capozzi:2025wyn,
    author = "Capozzi, Francesco and Giar{\`e}, William and Lisi, Eligio and Marrone, Antonio and Melchiorri, Alessandro and Palazzo, Antonio",
    title = "{Neutrino masses and mixing: Entering the era of subpercent precision}",
    eprint = "2503.07752",
    archivePrefix = "arXiv",
    primaryClass = "hep-ph",
    doi = "10.1103/PhysRevD.111.093006",
    journal = "Phys. Rev. D",
    volume = "111",
    number = "9",
    pages = "093006",
    year = "2025"
}

@article{Weinberg:1979sa,
    author = "Weinberg, Steven",
    title = "{Baryon and Lepton Nonconserving Processes}",
    reportNumber = "HUTP-79-A050",
    doi = "10.1103/PhysRevLett.43.1566",
    journal = "Phys. Rev. Lett.",
    volume = "43",
    pages = "1566--1570",
    year = "1979"
}

@article{Minkowski:1977sc,
    author = "Minkowski, Peter",
    title = "{$\mu \to e\gamma$ at a Rate of One Out of $10^{9}$ Muon Decays?}",
    reportNumber = "Print-77-0182 (BERN)",
    doi = "10.1016/0370-2693(77)90435-X",
    journal = "Phys. Lett. B",
    volume = "67",
    pages = "421--428",
    year = "1977"
}

@article{Mohapatra:1979ia,
    author = "Mohapatra, Rabindra N. and Senjanovic, Goran",
    title = "{Neutrino Mass and Spontaneous Parity Nonconservation}",
    reportNumber = "MDDP-TR-80-060, MDDP-PP-80-105, CCNY-HEP-79-10",
    doi = "10.1103/PhysRevLett.44.912",
    journal = "Phys. Rev. Lett.",
    volume = "44",
    pages = "912",
    year = "1980"
}

@article{Gell-Mann:1979vob,
    author = "Gell-Mann, Murray and Ramond, Pierre and Slansky, Richard",
    title = "{Complex Spinors and Unified Theories}",
    eprint = "1306.4669",
    archivePrefix = "arXiv",
    primaryClass = "hep-th",
    reportNumber = "PRINT-80-0576",
    journal = "Conf. Proc. C",
    volume = "790927",
    pages = "315--321",
    year = "1979"
}

@inbook{King:2025eqv,
    author = "King, Stephen F.",
    title = "{Right-handed neutrinos: seesaw models and signatures}",
    eprint = "2502.07877",
    archivePrefix = "arXiv",
    primaryClass = "hep-ph",
    month = "2",
    year = "2025"
}

@article{Deppisch:2012nb,
    author = "Deppisch, Frank F. and Hirsch, Martin and Pas, Heinrich",
    title = "{Neutrinoless Double Beta Decay and Physics Beyond the Standard Model}",
    eprint = "1208.0727",
    archivePrefix = "arXiv",
    primaryClass = "hep-ph",
    reportNumber = "IFIC-12-56",
    doi = "10.1088/0954-3899/39/12/124007",
    journal = "J. Phys. G",
    volume = "39",
    pages = "124007",
    year = "2012"
}

@article{DellOro:2016tmg,
    author = "Dell'Oro, Stefano and Marcocci, Simone and Viel, Matteo and Vissani, Francesco",
    title = "{Neutrinoless double beta decay: 2015 review}",
    eprint = "1601.07512",
    archivePrefix = "arXiv",
    primaryClass = "hep-ph",
    doi = "10.1155/2016/2162659",
    journal = "Adv. High Energy Phys.",
    volume = "2016",
    pages = "2162659",
    year = "2016"
}

@article{T2K:2023smv,
    author = "Abe, K. and others",
    collaboration = "T2K",
    title = "{Measurements of neutrino oscillation parameters from the T2K experiment using $3.6\times 10^{21}$ protons on target}",
    eprint = "2303.03222",
    archivePrefix = "arXiv",
    primaryClass = "hep-ex",
    doi = "10.1140/epjc/s10052-023-11819-x",
    journal = "Eur. Phys. J. C",
    volume = "83",
    number = "9",
    pages = "782",
    year = "2023"
}

@article{NOvA:2021nfi,
    author = "Acero, M. A. and others",
    collaboration = "NOvA",
    title = "{Improved measurement of neutrino oscillation parameters by the NOvA experiment}",
    eprint = "2108.08219",
    archivePrefix = "arXiv",
    primaryClass = "hep-ex",
    reportNumber = "FERMILAB-PUB-21-373-ND",
    doi = "10.1103/PhysRevD.106.032004",
    journal = "Phys. Rev. D",
    volume = "106",
    number = "3",
    pages = "032004",
    year = "2022"
}

@article{JUNO:2015zny,
    author = "An, Fengpeng and others",
    collaboration = "JUNO",
    title = "{Neutrino Physics with JUNO}",
    eprint = "1507.05613",
    archivePrefix = "arXiv",
    primaryClass = "physics.ins-det",
    doi = "10.1088/0954-3899/43/3/030401",
    journal = "J. Phys. G",
    volume = "43",
    number = "3",
    pages = "030401",
    year = "2016"
}

@article{Hyper-Kamiokande:2018ofw,
    author = "Abe, K. and others",
    collaboration = "Hyper-Kamiokande",
    title = "{Hyper-Kamiokande Design Report}",
    eprint = "1805.04163",
    archivePrefix = "arXiv",
    primaryClass = "physics.ins-det",
    month = "5",
    year = "2018"
}

@article{DUNE:2020ypp,
    author = "Abi, Babak and others",
    collaboration = "DUNE",
    title = "{Deep Underground Neutrino Experiment (DUNE), Far Detector Technical Design Report, Volume II: DUNE Physics}",
    eprint = "2002.03005",
    archivePrefix = "arXiv",
    primaryClass = "hep-ex",
    reportNumber = "FERMILAB-PUB-20-025-ND, FERMILAB-DESIGN-2020-02",
    month = "2",
    year = "2020"
}

@article{Gariazzo:2018pei,
    author = "Gariazzo, S. and Archidiacono, M. and de Salas, P. F. and Mena, O. and Ternes, C. A. and T{\'o}rtola, M.",
    title = "{Neutrino masses and their ordering: Global Data, Priors and Models}",
    eprint = "1801.04946",
    archivePrefix = "arXiv",
    primaryClass = "hep-ph",
    doi = "10.1088/1475-7516/2018/03/011",
    journal = "JCAP",
    volume = "03",
    pages = "011",
    year = "2018"
}

@article{Katrin:2024tvg,
    author = "Aker, Max and others",
    collaboration = "KATRIN",
    title = "{Direct neutrino-mass measurement based on 259 days of KATRIN data}",
    eprint = "2406.13516",
    archivePrefix = "arXiv",
    primaryClass = "nucl-ex",
    doi = "10.1126/science.adq9592",
    journal = "Science",
    volume = "388",
    number = "6743",
    pages = "adq9592",
    year = "2025"
}

@article{DESI:2025ejh,
    author = "Elbers, W. and others",
    collaboration = "DESI",
    title = "{Constraints on Neutrino Physics from DESI DR2 BAO and DR1 Full Shape}",
    eprint = "2503.14744",
    archivePrefix = "arXiv",
    primaryClass = "astro-ph.CO",
    reportNumber = "FERMILAB-PUB-25-0168-PPD",
    month = "3",
    year = "2025"
}

@article{Wang:2024hen,
    author = "Wang, Deng and Mena, Olga and Di Valentino, Eleonora and Gariazzo, Stefano",
    title = "{Updating neutrino mass constraints with background measurements}",
    eprint = "2405.03368",
    archivePrefix = "arXiv",
    primaryClass = "astro-ph.CO",
    doi = "10.1103/PhysRevD.110.103536",
    journal = "Phys. Rev. D",
    volume = "110",
    number = "10",
    pages = "103536",
    year = "2024"
}

@article{Jiang:2024viw,
    author = "Jiang, Jun-Qian and Giar{\`e}, William and Gariazzo, Stefano and Dainotti, Maria Giovanna and Di Valentino, Eleonora and Mena, Olga and Pedrotti, Davide and da Costa, Simony Santos and Vagnozzi, Sunny",
    title = "{Neutrino cosmology after DESI: tightest mass upper limits, preference for the normal ordering, and tension with terrestrial observations}",
    eprint = "2407.18047",
    archivePrefix = "arXiv",
    primaryClass = "astro-ph.CO",
    doi = "10.1088/1475-7516/2025/01/153",
    journal = "JCAP",
    volume = "01",
    pages = "153",
    year = "2025"
}

@article{Planck:2018vyg,
    author = "Aghanim, N. and others",
    collaboration = "Planck",
    title = "{Planck 2018 results. VI. Cosmological parameters}",
    eprint = "1807.06209",
    archivePrefix = "arXiv",
    primaryClass = "astro-ph.CO",
    doi = "10.1051/0004-6361/201833910",
    journal = "Astron. Astrophys.",
    volume = "641",
    pages = "A6",
    year = "2020",
    note = "[Erratum: Astron.Astrophys. 652, C4 (2021)]"
}

@article{Palanque-Delabrouille:2019iyz,
    author = {Palanque-Delabrouille, Nathalie and Y{\`e}che, Christophe and Sch{\"o}neberg, Nils and Lesgourgues, Julien and Walther, Michael and Chabanier, Sol{\`e}ne and Armengaud, Eric},
    title = "{Hints, neutrino bounds and WDM constraints from SDSS DR14 Lyman-$\alpha$ and Planck full-survey data}",
    eprint = "1911.09073",
    archivePrefix = "arXiv",
    primaryClass = "astro-ph.CO",
    doi = "10.1088/1475-7516/2020/04/038",
    journal = "JCAP",
    volume = "04",
    pages = "038",
    year = "2020"
}

@article{DiValentino:2021hoh,
    author = "Di Valentino, Eleonora and Gariazzo, Stefano and Mena, Olga",
    title = "{Most constraining cosmological neutrino mass bounds}",
    eprint = "2106.15267",
    archivePrefix = "arXiv",
    primaryClass = "astro-ph.CO",
    doi = "10.1103/PhysRevD.104.083504",
    journal = "Phys. Rev. D",
    volume = "104",
    number = "8",
    pages = "083504",
    year = "2021"
}

@article{Brieden:2022lsd,
    author = "Brieden, Samuel and Gil-Mar{\'\i}n, H{\'e}ctor and Verde, Licia",
    title = "{Model-agnostic interpretation of 10 billion years of cosmic evolution traced by BOSS and eBOSS data}",
    eprint = "2204.11868",
    archivePrefix = "arXiv",
    primaryClass = "astro-ph.CO",
    doi = "10.1088/1475-7516/2022/08/024",
    journal = "JCAP",
    volume = "08",
    number = "08",
    pages = "024",
    year = "2022"
}

@article{Craig:2024tky,
    author = "Craig, Nathaniel and Green, Daniel and Meyers, Joel and Rajendran, Surjeet",
    title = "{No {\ensuremath{\nu}}s is Good News}",
    eprint = "2405.00836",
    archivePrefix = "arXiv",
    primaryClass = "astro-ph.CO",
    reportNumber = "FERMILAB-PUB-24-0492-SQMS-V",
    doi = "10.1007/JHEP09(2024)097",
    journal = "JHEP",
    volume = "09",
    pages = "097",
    year = "2024"
}

@article{Naredo-Tuero:2024sgf,
    author = "Naredo-Tuero, Daniel and Escudero, Miguel and Fern{\'a}ndez-Mart{\'\i}nez, Enrique and Marcano, Xabier and Poulin, Vivian",
    title = "{Critical look at the cosmological neutrino mass bound}",
    eprint = "2407.13831",
    archivePrefix = "arXiv",
    primaryClass = "astro-ph.CO",
    reportNumber = "CERN-TH-2024-115, IFT-UAM/CSIC-24-106",
    doi = "10.1103/PhysRevD.110.123537",
    journal = "Phys. Rev. D",
    volume = "110",
    number = "12",
    pages = "123537",
    year = "2024"
}

@article{Green:2024xbb,
    author = "Green, Daniel and Meyers, Joel",
    title = "{Cosmological preference for a negative neutrino mass}",
    eprint = "2407.07878",
    archivePrefix = "arXiv",
    primaryClass = "astro-ph.CO",
    doi = "10.1103/PhysRevD.111.083507",
    journal = "Phys. Rev. D",
    volume = "111",
    number = "8",
    pages = "083507",
    year = "2025"
}

@article{Elbers:2024sha,
    author = "Elbers, Willem and Frenk, Carlos S. and Jenkins, Adrian and Li, Baojiu and Pascoli, Silvia",
    title = "{Negative neutrino masses as a mirage of dark energy}",
    eprint = "2407.10965",
    archivePrefix = "arXiv",
    primaryClass = "astro-ph.CO",
    doi = "10.1103/PhysRevD.111.063534",
    journal = "Phys. Rev. D",
    volume = "111",
    number = "6",
    pages = "063534",
    year = "2025"
}

@article{Colgain:2024mtg,
    author = "Colg{\'a}in, Eoin {\'O}. and Sheikh-Jabbari, M. M.",
    title = "{DESI and SNe: Dynamical Dark Energy, $\Omega_m$ Tension or Systematics?}",
    eprint = "2412.12905",
    archivePrefix = "arXiv",
    primaryClass = "astro-ph.CO",
    month = "12",
    year = "2024"
}

@article{Lynch:2025ine,
    author = "Lynch, Gabriel P. and Knox, Lloyd",
    title = "{What's the matter with $\Sigma m_{\nu}$?}",
    eprint = "2503.14470",
    archivePrefix = "arXiv",
    primaryClass = "astro-ph.CO",
    month = "3",
    year = "2025"
}

@article{Sailer:2025lxj,
    author = "Sailer, Noah and Farren, Gerrit S. and Ferraro, Simone and White, Martin",
    title = "{Dispu$\tau$able: the high cost of a low optical depth}",
    eprint = "2504.16932",
    archivePrefix = "arXiv",
    primaryClass = "astro-ph.CO",
    month = "4",
    year = "2025"
}

@article{Jhaveri:2025neg,
    author = "Jhaveri, Tanisha and Karwal, Tanvi and Hu, Wayne",
    title = "{Turning a negative neutrino mass into a positive optical depth}",
    eprint = "2504.21813",
    archivePrefix = "arXiv",
    primaryClass = "astro-ph.CO",
    month = "4",
    year = "2025"
}

@article{Wang:2024pui,
    author = "Wang, Zhengyi and Lin, Shijie and Ding, Zhejie and Hu, Bin",
    title = "{The role of LRG1 and LRG2{\textquoteright}s monopole in inferring the DESI 2024 BAO cosmology}",
    eprint = "2405.02168",
    archivePrefix = "arXiv",
    primaryClass = "astro-ph.CO",
    doi = "10.1093/mnras/stae2309",
    journal = "Mon. Not. Roy. Astron. Soc.",
    volume = "534",
    number = "4",
    pages = "3869--3875",
    year = "2024"
}

@article{Colgain:2024xqj,
    author = "Colg{\'a}in, Eoin {\'O}. and Dainotti, Maria Giovanna and Capozziello, Salvatore and Pourojaghi, Saeed and Sheikh-Jabbari, M. M. and Stojkovic, Dejan",
    title = "{Does DESI 2024 Confirm $\Lambda$CDM?}",
    eprint = "2404.08633",
    archivePrefix = "arXiv",
    primaryClass = "astro-ph.CO",
    month = "4",
    year = "2024"
}

@article{Sapone:2024ltl,
    author = "Sapone, Domenico and Nesseris, Savvas",
    title = "{Outliers in DESI BAO: robustness and cosmological implications}",
    eprint = "2412.01740",
    archivePrefix = "arXiv",
    primaryClass = "astro-ph.CO",
    reportNumber = "IFT-UAM/CSIC-24-169",
    month = "12",
    year = "2024"
}

@article{Colgain:2025fct,
    author = "Colg{\'a}in, Eoin {\'O}. and Pourojaghi, Saeed and Sheikh-Jabbari, M. M.",
    title = "{On the Pipeline Dependence of DESI Dynamical Dark Energy}",
    eprint = "2505.19029",
    archivePrefix = "arXiv",
    primaryClass = "astro-ph.CO",
    month = "5",
    year = "2025"
}

@article{Colgain:2025nzf,
    author = "Colg{\'a}in, Eoin {\'O}. and Pourojaghi, Saeed and Sheikh-Jabbari, M. M. and Yin, Lu",
    title = "{How much has DESI dark energy evolved since DR1?}",
    eprint = "2504.04417",
    archivePrefix = "arXiv",
    primaryClass = "astro-ph.CO",
    month = "4",
    year = "2025"
}

@article{Giare:2025pzu,
    author = "Giar{\`e}, William and Mahassen, Tariq and Di Valentino, Eleonora and Pan, Supriya",
    title = "{An overview of what current data can (and cannot yet) say about evolving dark energy}",
    eprint = "2502.10264",
    archivePrefix = "arXiv",
    primaryClass = "astro-ph.CO",
    doi = "10.1016/j.dark.2025.101906",
    journal = "Phys. Dark Univ.",
    volume = "48",
    pages = "101906",
    year = "2025"
}

@article{Efstathiou:2025tie,
    author = "Efstathiou, George",
    title = "{Baryon Acoustic Oscillations from a Different Angle}",
    eprint = "2505.02658",
    archivePrefix = "arXiv",
    primaryClass = "astro-ph.CO",
    month = "5",
    year = "2025"
}

@article{Giare:2024oil,
    author = "Giar{\`e}, William",
    title = "{Dynamical Dark Energy Beyond Planck? Constraints from multiple CMB probes, DESI BAO and Type-Ia Supernovae}",
    eprint = "2409.17074",
    archivePrefix = "arXiv",
    primaryClass = "astro-ph.CO",
    month = "9",
    year = "2024"
}

@article{Giare:2023ejv,
    author = "Giar{\`e}, William and Di Valentino, Eleonora and Melchiorri, Alessandro",
    title = "{Measuring the reionization optical depth without large-scale CMB polarization}",
    eprint = "2312.06482",
    archivePrefix = "arXiv",
    primaryClass = "astro-ph.CO",
    doi = "10.1103/PhysRevD.109.103519",
    journal = "Phys. Rev. D",
    volume = "109",
    number = "10",
    pages = "103519",
    year = "2024"
}

@article{RoyChoudhury:2024wri,
    author = "Roy Choudhury, Shouvik and Okumura, Teppei",
    title = "{Updated Cosmological Constraints in Extended Parameter Space with Planck PR4, DESI Baryon Acoustic Oscillations, and Supernovae: Dynamical Dark Energy, Neutrino Masses, Lensing Anomaly, and the Hubble Tension}",
    eprint = "2409.13022",
    archivePrefix = "arXiv",
    primaryClass = "astro-ph.CO",
    doi = "10.3847/2041-8213/ad8c26",
    journal = "Astrophys. J. Lett.",
    volume = "976",
    number = "1",
    pages = "L11",
    year = "2024"
}

@article{Gariazzo:2024sil,
    author = "Gariazzo, Stefano and Giar{\`e}, William and Mena, Olga and Di Valentino, Eleonora",
    title = "{How robust are the parameter constraints extending the {\ensuremath{\Lambda}}CDM model?}",
    eprint = "2404.11182",
    archivePrefix = "arXiv",
    primaryClass = "astro-ph.CO",
    doi = "10.1103/PhysRevD.111.023540",
    journal = "Phys. Rev. D",
    volume = "111",
    number = "2",
    pages = "023540",
    year = "2025"
}

@article{RoyChoudhury:2025dhe,
    author = "Roy Choudhury, Shouvik",
    title = "{Cosmology in Extended Parameter Space with DESI Data Release 2 Baryon Acoustic Oscillations: A 2$σ$+ Detection of Nonzero Neutrino Masses with an Update on Dynamical Dark Energy and Lensing Anomaly}",
    eprint = "2504.15340",
    archivePrefix = "arXiv",
    primaryClass = "astro-ph.CO",
    doi = "10.3847/2041-8213/ade1cc",
    journal = "Astrophys. J. Lett.",
    volume = "986",
    pages = "L31",
    month = "4",
    year = "2025"
}

@article{DESI:2024mwx,
    author = "Adame, A. G. and others",
    collaboration = "DESI",
    title = "{DESI 2024 VI: cosmological constraints from the measurements of baryon acoustic oscillations}",
    eprint = "2404.03002",
    archivePrefix = "arXiv",
    primaryClass = "astro-ph.CO",
    reportNumber = "FERMILAB-PUB-24-0154-PPD",
    doi = "10.1088/1475-7516/2025/02/021",
    journal = "JCAP",
    volume = "02",
    pages = "021",
    year = "2025"
}

@article{DESI:2025zgx,
    author = "Abdul Karim, M. and others",
    collaboration = "DESI",
    title = "{DESI DR2 Results II: Measurements of Baryon Acoustic Oscillations and Cosmological Constraints}",
    eprint = "2503.14738",
    archivePrefix = "arXiv",
    primaryClass = "astro-ph.CO",
    reportNumber = "FERMILAB-PUB-25-0169-PPD",
    month = "3",
    year = "2025"
}

@article{DESI:2025wyn,
    author = "Gu, Gan and others",
    collaboration = "DESI",
    title = "{Dynamical Dark Energy in light of the DESI DR2 Baryonic Acoustic Oscillations Measurements}",
    eprint = "2504.06118",
    archivePrefix = "arXiv",
    primaryClass = "astro-ph.CO",
    reportNumber = "FERMILAB-PUB-25-0235-PPD",
    month = "4",
    year = "2025"
}

@article{Cortes:2024lgw,
    author = "Cort{\^e}s, Marina and Liddle, Andrew R.",
    title = "{Interpreting DESI's evidence for evolving dark energy}",
    eprint = "2404.08056",
    archivePrefix = "arXiv",
    primaryClass = "astro-ph.CO",
    doi = "10.1088/1475-7516/2024/12/007",
    journal = "JCAP",
    volume = "12",
    pages = "007",
    year = "2024"
}

@article{Shlivko:2024llw,
    author = "Shlivko, David and Steinhardt, Paul J.",
    title = "{Assessing observational constraints on dark energy}",
    eprint = "2405.03933",
    archivePrefix = "arXiv",
    primaryClass = "astro-ph.CO",
    doi = "10.1016/j.physletb.2024.138826",
    journal = "Phys. Lett. B",
    volume = "855",
    pages = "138826",
    year = "2024"
}

@article{Luongo:2024fww,
    author = "Luongo, Orlando and Muccino, Marco",
    title = "{Model-independent cosmographic constraints from DESI 2024}",
    eprint = "2404.07070",
    archivePrefix = "arXiv",
    primaryClass = "astro-ph.CO",
    doi = "10.1051/0004-6361/202450512",
    journal = "Astron. Astrophys.",
    volume = "690",
    pages = "A40",
    year = "2024"
}

@article{Yin:2024hba,
    author = "Yin, Wen",
    title = "{Cosmic clues: DESI, dark energy, and the cosmological constant problem}",
    eprint = "2404.06444",
    archivePrefix = "arXiv",
    primaryClass = "hep-ph",
    doi = "10.1007/JHEP05(2024)327",
    journal = "JHEP",
    volume = "05",
    pages = "327",
    year = "2024"
}

@article{Gialamas:2024lyw,
    author = {Gialamas, Ioannis D. and H{\"u}tsi, Gert and Kannike, Kristjan and Racioppi, Antonio and Raidal, Martti and Vasar, Martin and Veerm{\"a}e, Hardi},
    title = "{Interpreting DESI 2024 BAO: Late-time dynamical dark energy or a local effect?}",
    eprint = "2406.07533",
    archivePrefix = "arXiv",
    primaryClass = "astro-ph.CO",
    doi = "10.1103/PhysRevD.111.043540",
    journal = "Phys. Rev. D",
    volume = "111",
    number = "4",
    pages = "043540",
    year = "2025"
}

@article{Dinda:2024kjf,
    author = "Dinda, Bikash R.",
    title = "{A new diagnostic for the null test of dynamical dark energy in light of DESI 2024 and other BAO data}",
    eprint = "2405.06618",
    archivePrefix = "arXiv",
    primaryClass = "astro-ph.CO",
    doi = "10.1088/1475-7516/2024/09/062",
    journal = "JCAP",
    volume = "09",
    pages = "062",
    year = "2024"
}

@article{Najafi:2024qzm,
    author = "Najafi, Mahdi and Pan, Supriya and Di Valentino, Eleonora and Firouzjaee, Javad T.",
    title = "{Dynamical dark energy confronted with multiple CMB missions}",
    eprint = "2407.14939",
    archivePrefix = "arXiv",
    primaryClass = "astro-ph.CO",
    doi = "10.1016/j.dark.2024.101539",
    journal = "Phys. Dark Univ.",
    volume = "45",
    pages = "101539",
    year = "2024"
}

@article{Wang:2024dka,
    author = "Wang, Hao and Piao, Yun-Song",
    title = "{Dark energy in light of recent DESI BAO and Hubble tension}",
    eprint = "2404.18579",
    archivePrefix = "arXiv",
    primaryClass = "astro-ph.CO",
    month = "4",
    year = "2024"
}

@article{Ye:2024ywg,
    author = "Ye, Gen and Martinelli, Matteo and Hu, Bin and Silvestri, Alessandra",
    title = "{Hints of Nonminimally Coupled Gravity in DESI 2024 Baryon Acoustic Oscillation Measurements}",
    eprint = "2407.15832",
    archivePrefix = "arXiv",
    primaryClass = "astro-ph.CO",
    doi = "10.1103/PhysRevLett.134.181002",
    journal = "Phys. Rev. Lett.",
    volume = "134",
    number = "18",
    pages = "181002",
    year = "2025"
}

@article{Tada:2024znt,
    author = "Tada, Yuichiro and Terada, Takahiro",
    title = "{Quintessential interpretation of the evolving dark energy in light of DESI observations}",
    eprint = "2404.05722",
    archivePrefix = "arXiv",
    primaryClass = "astro-ph.CO",
    doi = "10.1103/PhysRevD.109.L121305",
    journal = "Phys. Rev. D",
    volume = "109",
    number = "12",
    pages = "L121305",
    year = "2024"
}

@article{Carloni:2024zpl,
    author = "Carloni, Youri and Luongo, Orlando and Muccino, Marco",
    title = "{Does dark energy really revive using DESI 2024 data?}",
    eprint = "2404.12068",
    archivePrefix = "arXiv",
    primaryClass = "astro-ph.CO",
    doi = "10.1103/PhysRevD.111.023512",
    journal = "Phys. Rev. D",
    volume = "111",
    number = "2",
    pages = "023512",
    year = "2025"
}

@article{Chan-GyungPark:2024mlx,
    author = "Park, Chan-Gyung and de Cruz P{\'e}rez, Javier and Ratra, Bharat",
    title = "{Using non-DESI data to confirm and strengthen the DESI 2024 spatially flat w0waCDM cosmological parametrization result}",
    eprint = "2405.00502",
    archivePrefix = "arXiv",
    primaryClass = "astro-ph.CO",
    doi = "10.1103/PhysRevD.110.123533",
    journal = "Phys. Rev. D",
    volume = "110",
    number = "12",
    pages = "123533",
    year = "2024"
}

@article{DESI:2024kob,
    author = "Lodha, K. and others",
    collaboration = "DESI",
    title = "{DESI 2024: Constraints on physics-focused aspects of dark energy using DESI DR1 BAO data}",
    eprint = "2405.13588",
    archivePrefix = "arXiv",
    primaryClass = "astro-ph.CO",
    reportNumber = "FERMILAB-PUB-24-0756-PPD",
    doi = "10.1103/PhysRevD.111.023532",
    journal = "Phys. Rev. D",
    volume = "111",
    number = "2",
    pages = "023532",
    year = "2025"
}

@article{Ramadan:2024kmn,
    author = "Ramadan, Omar F. and Sakstein, Jeremy and Rubin, David",
    title = "{DESI constraints on exponential quintessence}",
    eprint = "2405.18747",
    archivePrefix = "arXiv",
    primaryClass = "astro-ph.CO",
    doi = "10.1103/PhysRevD.110.L041303",
    journal = "Phys. Rev. D",
    volume = "110",
    number = "4",
    pages = "L041303",
    year = "2024"
}

@article{Notari:2024rti,
    author = "Notari, Alessio and Redi, Michele and Tesi, Andrea",
    title = "{Consistent theories for the DESI dark energy fit}",
    eprint = "2406.08459",
    archivePrefix = "arXiv",
    primaryClass = "astro-ph.CO",
    doi = "10.1088/1475-7516/2024/11/025",
    journal = "JCAP",
    volume = "11",
    pages = "025",
    year = "2024"
}

@article{Orchard:2024bve,
    author = "Orchard, Lili and C{\'a}rdenas, V{\'\i}ctor H.",
    title = "{Probing dark energy evolution post-DESI 2024}",
    eprint = "2407.05579",
    archivePrefix = "arXiv",
    primaryClass = "astro-ph.CO",
    doi = "10.1016/j.dark.2024.101678",
    journal = "Phys. Dark Univ.",
    volume = "46",
    pages = "101678",
    year = "2024"
}

@article{Hernandez-Almada:2024ost,
    author = "Hern{\'a}ndez-Almada, A. and Mendoza-Mart{\'\i}nez, M. L. and Garc{\'\i}a-Aspeitia, Miguel A. and Motta, V.",
    title = "{Phenomenological emergent dark energy in the light of DESI Data Release 1}",
    eprint = "2407.09430",
    archivePrefix = "arXiv",
    primaryClass = "astro-ph.CO",
    doi = "10.1016/j.dark.2024.101668",
    journal = "Phys. Dark Univ.",
    volume = "46",
    pages = "101668",
    year = "2024"
}

@article{Pourojaghi:2024tmw,
    author = "Malekjani, Mohammad and Davari, Zahra and Pourojaghi, Saeed",
    collaboration = "DESI",
    title = "{Cosmological constraints on dark energy parametrizations after DESI 2024: Persistent deviation from standard {\ensuremath{\Lambda}}CDM cosmology}",
    eprint = "2407.09767",
    archivePrefix = "arXiv",
    primaryClass = "astro-ph.CO",
    doi = "10.1103/PhysRevD.111.083547",
    journal = "Phys. Rev. D",
    volume = "111",
    number = "8",
    pages = "083547",
    year = "2025"
}

@article{Giare:2024gpk,
    author = "Giar{\`e}, William and Najafi, Mahdi and Pan, Supriya and Di Valentino, Eleonora and Firouzjaee, Javad T.",
    title = "{Robust preference for Dynamical Dark Energy in DESI BAO and SN measurements}",
    eprint = "2407.16689",
    archivePrefix = "arXiv",
    primaryClass = "astro-ph.CO",
    doi = "10.1088/1475-7516/2024/10/035",
    journal = "JCAP",
    volume = "10",
    pages = "035",
    year = "2024"
}

@article{Reboucas:2024smm,
    author = "Rebou{\c{c}}as, Jo{\~a}o and de Souza, Diogo H. F. and Zhong, Kunhao and Miranda, Vivian and Rosenfeld, Rogerio",
    title = "{Investigating late-time dark energy and massive neutrinos in light of DESI Y1 BAO}",
    eprint = "2408.14628",
    archivePrefix = "arXiv",
    primaryClass = "astro-ph.CO",
    doi = "10.1088/1475-7516/2025/02/024",
    journal = "JCAP",
    volume = "02",
    pages = "024",
    year = "2025"
}

@article{Giare:2024ocw,
    author = "Giar{\`e}, William",
    title = "{Dynamical Dark Energy Beyond Planck? Constraints from multiple CMB probes, DESI BAO and Type-Ia Supernovae}",
    eprint = "2409.17074",
    archivePrefix = "arXiv",
    primaryClass = "astro-ph.CO",
    month = "9",
    year = "2024"
}

@article{Chan-GyungPark:2024brx,
    author = "Park, Chan-Gyung and de Cruz P{\'e}rez, Javier and Ratra, Bharat",
    title = "{Is the $w_0w_a$CDM cosmological parameterization evidence for dark energy dynamics partially caused by the excess smoothing of Planck CMB anisotropy data?}",
    eprint = "2410.13627",
    archivePrefix = "arXiv",
    primaryClass = "astro-ph.CO",
    month = "10",
    year = "2024"
}

@article{Menci:2024hop,
    author = "Menci, Nicola and Sen, Anjan Ananda and Castellano, Marco",
    title = "{The Excess of JWST Bright Galaxies: A Possible Origin in the Ground State of Dynamical Dark Energy in the Light of DESI 2024 Data}",
    eprint = "2410.22940",
    archivePrefix = "arXiv",
    primaryClass = "astro-ph.CO",
    doi = "10.3847/1538-4357/ad8d5b",
    journal = "Astrophys. J.",
    volume = "976",
    number = "2",
    pages = "227",
    year = "2024"
}

@article{Li:2024qus,
    author = "Li, Tian-Nuo and Li, Yun-He and Du, Guo-Hong and Wu, Peng-Ju and Feng, Lu and Zhang, Jing-Fei and Zhang, Xin",
    title = "{Revisiting holographic dark energy after DESI 2024}",
    eprint = "2411.08639",
    archivePrefix = "arXiv",
    primaryClass = "astro-ph.CO",
    doi = "10.1140/epjc/s10052-025-14279-7",
    journal = "Eur. Phys. J. C",
    volume = "85",
    number = "6",
    pages = "608",
    year = "2025"
}

@article{Li:2024hrv,
    author = "Li, Jun-Xian and Wang, Shuang",
    title = "{A comprehensive numerical study on four categories of holographic dark energy models}",
    eprint = "2412.09064",
    archivePrefix = "arXiv",
    primaryClass = "astro-ph.CO",
    month = "12",
    year = "2024"
}

@article{Notari:2024zmi,
    author = "Notari, Alessio and Redi, Michele and Tesi, Andrea",
    title = "{BAO vs. SN evidence for evolving dark energy}",
    eprint = "2411.11685",
    archivePrefix = "arXiv",
    primaryClass = "astro-ph.CO",
    doi = "10.1088/1475-7516/2025/04/048",
    journal = "JCAP",
    volume = "04",
    pages = "048",
    year = "2025"
}

@article{Gao:2024ily,
    author = "Gao, Qing and Peng, Zhiqian and Gao, Shengqing and Gong, Yungui",
    title = "{On the Evidence of Dynamical Dark Energy}",
    eprint = "2411.16046",
    archivePrefix = "arXiv",
    primaryClass = "astro-ph.CO",
    doi = "10.3390/universe11010010",
    journal = "Universe",
    volume = "11",
    number = "1",
    pages = "10",
    year = "2025"
}

@article{Fikri:2024klc,
    author = "Fikri, Ramy and ElKhateeb, Esraa and Lashin, El Sayed and El Hanafy, Waleed",
    title = "{A preference for dynamical phantom dark energy using one-parameter model with Planck, DESI DR1 BAO and SN data}",
    eprint = "2411.19362",
    archivePrefix = "arXiv",
    primaryClass = "astro-ph.CO",
    month = "11",
    year = "2024"
}

@article{Jiang:2024xnu,
    author = "Jiang, Jun-Qian and Pedrotti, Davide and da Costa, Simony Santos and Vagnozzi, Sunny",
    title = "{Nonparametric late-time expansion history reconstruction and implications for the Hubble tension in light of recent DESI and type Ia supernovae data}",
    eprint = "2408.02365",
    archivePrefix = "arXiv",
    primaryClass = "astro-ph.CO",
    doi = "10.1103/PhysRevD.110.123519",
    journal = "Phys. Rev. D",
    volume = "110",
    number = "12",
    pages = "123519",
    year = "2024"
}

@article{Zheng:2024qzi,
    author = "Zheng, Jie and Qiang, Da-Chun and You, Zhi-Qiang",
    title = "{Cosmological constraints on dark energy models using DESI BAO 2024}",
    eprint = "2412.04830",
    archivePrefix = "arXiv",
    primaryClass = "astro-ph.CO",
    month = "12",
    year = "2024"
}

@article{Gomez-Valent:2024ejh,
    author = "G{\'o}mez-Valent, Adria and Sol{\`a} Peracaula, Joan",
    title = "{Composite dark energy and the cosmological tensions}",
    eprint = "2412.15124",
    archivePrefix = "arXiv",
    primaryClass = "astro-ph.CO",
    doi = "10.1016/j.physletb.2025.139391",
    journal = "Phys. Lett. B",
    volume = "864",
    pages = "139391",
    year = "2025"
}

@article{Lewis:2024cqj,
    author = "Lewis, Antony and Chamberlain, Ewan",
    title = "{Understanding acoustic scale observations: the one-sided fight against {\ensuremath{\Lambda}}}",
    eprint = "2412.13894",
    archivePrefix = "arXiv",
    primaryClass = "astro-ph.CO",
    doi = "10.1088/1475-7516/2025/05/065",
    journal = "JCAP",
    volume = "05",
    pages = "065",
    year = "2025"
}

@article{Wolf:2025jlc,
    author = "Wolf, William J. and Garc{\'\i}a-Garc{\'\i}a, Carlos and Ferreira, Pedro G.",
    title = "{Robustness of dark energy phenomenology across different parameterizations}",
    eprint = "2502.04929",
    archivePrefix = "arXiv",
    primaryClass = "astro-ph.CO",
    doi = "10.1088/1475-7516/2025/05/034",
    journal = "JCAP",
    volume = "05",
    pages = "034",
    year = "2025"
}

@article{Shajib:2025tpd,
    author = "Shajib, Anowar J. and Frieman, Joshua A.",
    title = "{Evolving dark energy models: Current and forecast constraints}",
    eprint = "2502.06929",
    archivePrefix = "arXiv",
    primaryClass = "astro-ph.CO",
    month = "2",
    year = "2025"
}

@article{Chaussidon:2025npr,
    author = "Chaussidon, E. and others",
    title = "{Early time solution as an alternative to the late time evolving dark energy with DESI DR2 BAO}",
    eprint = "2503.24343",
    archivePrefix = "arXiv",
    primaryClass = "astro-ph.CO",
    reportNumber = "FERMILAB-PUB-25-0241-PPD",
    month = "3",
    year = "2025"
}

@article{Kessler:2025kju,
    author = "Kessler, Daniel A. and Escamilla, Luis A. and Pan, Supriya and Di Valentino, Eleonora",
    title = "{One-parameter dynamical dark energy: Hints for oscillations}",
    eprint = "2504.00776",
    archivePrefix = "arXiv",
    primaryClass = "astro-ph.CO",
    month = "4",
    year = "2025"
}

@article{Pang:2025lvh,
    author = "Pang, Ye-Huang and Zhang, Xue and Huang, Qing-Guo",
    title = "{The impact of the Hubble tension on the evidence for dynamical dark energy}",
    eprint = "2503.21600",
    archivePrefix = "arXiv",
    primaryClass = "astro-ph.CO",
    doi = "10.1007/s11433-025-2713-8",
    journal = "Sci. China Phys. Mech. Astron.",
    volume = "68",
    number = "8",
    pages = "280410",
    year = "2025"
}

@article{Scherer:2025esj,
    author = "Scherer, Mateus and Sabogal, Miguel A. and Nunes, Rafael C. and De Felice, Antonio",
    title = "{Challenging $\Lambda$CDM: 5$\sigma$ Evidence for a Dynamical Dark Energy Late-Time Transition}",
    eprint = "2504.20664",
    archivePrefix = "arXiv",
    primaryClass = "astro-ph.CO",
    month = "4",
    year = "2025"
}

@article{Specogna:2025guo,
    author = "Specogna, Enrico and Adil, Shahnawaz A. and Ozulker, Emre and Di Valentino, Eleonora and Nunes, Rafael C. and Akarsu, Ozgur and Sen, Anjan A.",
    title = "{Updated Constraints on Omnipotent Dark Energy: A Comprehensive Analysis with CMB and BAO Data}",
    eprint = "2504.17859",
    archivePrefix = "arXiv",
    primaryClass = "gr-qc",
    month = "4",
    year = "2025"
}

@article{Teixeira:2025czm,
    author = "Teixeira, Elsa M. and Giar{\`e}, William and Hogg, Natalie B. and Montandon, Thomas and Poudou, Ad{\`e}le and Poulin, Vivian",
    title = "{Implications of distance duality violation for the $H_0$ tension and evolving dark energy}",
    eprint = "2504.10464",
    archivePrefix = "arXiv",
    primaryClass = "astro-ph.CO",
    month = "4",
    year = "2025"
}

@article{Cheng:2025lod,
    author = "Cheng, Hanyu and Di Valentino, Eleonora and Escamilla, Luis A. and Sen, Anjan A. and Visinelli, Luca",
    title = "{Pressure Parametrization of Dark Energy: First and Second-Order Constraints with Latest Cosmological Data}",
    eprint = "2505.02932",
    archivePrefix = "arXiv",
    primaryClass = "astro-ph.CO",
    month = "5",
    year = "2025"
}

@article{Cheng:2025hug,
    author = "Cheng, Hanyu and Di Valentino, Eleonora and Visinelli, Luca",
    title = "{Cosmic Strings as Dynamical Dark Energy: Novel Constraints}",
    eprint = "2505.22066",
    archivePrefix = "arXiv",
    primaryClass = "astro-ph.CO",
    month = "5",
    year = "2025"
}

@article{Ozulker:2025ehg,
    author = {{\"O}z{\"u}lker, Emre and Di Valentino, Eleonora and Giar{\`e}, William},
    title = "{Dark Energy Crosses the Line: Quantifying and Testing the Evidence for Phantom Crossing}",
    eprint = "2506.19053",
    archivePrefix = "arXiv",
    primaryClass = "astro-ph.CO",
    month = "6",
    year = "2025"
}

@article{Chevallier:2000qy,
    author = "Chevallier, Michel and Polarski, David",
    title = "{Accelerating universes with scaling dark matter}",
    eprint = "gr-qc/0009008",
    archivePrefix = "arXiv",
    doi = "10.1142/S0218271801000822",
    journal = "Int. J. Mod. Phys. D",
    volume = "10",
    pages = "213--224",
    year = "2001"
}

@article{Linder:2002et,
    author = "Linder, Eric V.",
    title = "{Exploring the expansion history of the universe}",
    eprint = "astro-ph/0208512",
    archivePrefix = "arXiv",
    doi = "10.1103/PhysRevLett.90.091301",
    journal = "Phys. Rev. Lett.",
    volume = "90",
    pages = "091301",
    year = "2003"
}

@article{Efstathiou:2024xcq,
    author = "Efstathiou, George",
    title = "{Evolving dark energy or supernovae systematics?}",
    eprint = "2408.07175",
    archivePrefix = "arXiv",
    primaryClass = "astro-ph.CO",
    doi = "10.1093/mnras/staf301",
    journal = "Mon. Not. Roy. Astron. Soc.",
    volume = "538",
    number = "2",
    pages = "875--882",
    year = "2025"
}

@article{Ye:2025ark,
    author = "Ye, Gen and Lin, Shi-Jie",
    title = "{On the tension between DESI DR2 BAO and CMB}",
    eprint = "2505.02207",
    archivePrefix = "arXiv",
    primaryClass = "astro-ph.CO",
    month = "5",
    year = "2025"
}

@article{Verde:2019ivm,
    author = "Verde, L. and Treu, T. and Riess, A. G.",
    title = "{Tensions between the Early and the Late Universe}",
    eprint = "1907.10625",
    archivePrefix = "arXiv",
    primaryClass = "astro-ph.CO",
    doi = "10.1038/s41550-019-0902-0",
    journal = "Nature Astron.",
    volume = "3",
    pages = "891",
    year = "2019"
}

@article{DiValentino:2020zio,
    author = "Di Valentino, Eleonora and others",
    title = "{Snowmass2021 - Letter of interest cosmology intertwined II: The hubble constant tension}",
    eprint = "2008.11284",
    archivePrefix = "arXiv",
    primaryClass = "astro-ph.CO",
    reportNumber = "FERMILAB-PUB-21-590-PPD",
    doi = "10.1016/j.astropartphys.2021.102605",
    journal = "Astropart. Phys.",
    volume = "131",
    pages = "102605",
    year = "2021"
}

@article{DiValentino:2021izs,
    author = "Di Valentino, Eleonora and Mena, Olga and Pan, Supriya and Visinelli, Luca and Yang, Weiqiang and Melchiorri, Alessandro and Mota, David F. and Riess, Adam G. and Silk, Joseph",
    title = "{In the realm of the Hubble tension{\textemdash}a review of solutions}",
    eprint = "2103.01183",
    archivePrefix = "arXiv",
    primaryClass = "astro-ph.CO",
    reportNumber = "IPPP/20/108",
    doi = "10.1088/1361-6382/ac086d",
    journal = "Class. Quant. Grav.",
    volume = "38",
    number = "15",
    pages = "153001",
    year = "2021"
}

@article{Perivolaropoulos:2021jda,
    author = "Perivolaropoulos, Leandros and Skara, Foteini",
    title = "{Challenges for {\ensuremath{\Lambda}}CDM: An update}",
    eprint = "2105.05208",
    archivePrefix = "arXiv",
    primaryClass = "astro-ph.CO",
    doi = "10.1016/j.newar.2022.101659",
    journal = "New Astron. Rev.",
    volume = "95",
    pages = "101659",
    year = "2022"
}

@article{Schoneberg:2021qvd,
    author = {Sch{\"o}neberg, Nils and Franco Abell{\'a}n, Guillermo and P{\'e}rez S{\'a}nchez, Andrea and Witte, Samuel J. and Poulin, Vivian and Lesgourgues, Julien},
    title = "{The H0 Olympics: A fair ranking of proposed models}",
    eprint = "2107.10291",
    archivePrefix = "arXiv",
    primaryClass = "astro-ph.CO",
    doi = "10.1016/j.physrep.2022.07.001",
    journal = "Phys. Rept.",
    volume = "984",
    pages = "1--55",
    year = "2022"
}

@article{Shah:2021onj,
    author = "Shah, Paul and Lemos, Pablo and Lahav, Ofer",
    title = "{A buyer{\textquoteright}s guide to the Hubble constant}",
    eprint = "2109.01161",
    archivePrefix = "arXiv",
    primaryClass = "astro-ph.CO",
    doi = "10.1007/s00159-021-00137-4",
    journal = "Astron. Astrophys. Rev.",
    volume = "29",
    number = "1",
    pages = "9",
    year = "2021"
}

@article{Abdalla:2022yfr,
    author = "Abdalla, Elcio and others",
    title = "{Cosmology intertwined: A review of the particle physics, astrophysics, and cosmology associated with the cosmological tensions and anomalies}",
    eprint = "2203.06142",
    archivePrefix = "arXiv",
    primaryClass = "astro-ph.CO",
    reportNumber = "FERMILAB-CONF-22-192-SCD",
    doi = "10.1016/j.jheap.2022.04.002",
    journal = "JHEAp",
    volume = "34",
    pages = "49--211",
    year = "2022"
}

@article{DiValentino:2022fjm,
    author = "Di Valentino, Eleonora",
    title = "{Challenges of the Standard Cosmological Model}",
    doi = "10.3390/universe8080399",
    journal = "Universe",
    volume = "8",
    number = "8",
    pages = "399",
    year = "2022"
}

@article{Kamionkowski:2022pkx,
    author = "Kamionkowski, Marc and Riess, Adam G.",
    title = "{The Hubble Tension and Early Dark Energy}",
    eprint = "2211.04492",
    archivePrefix = "arXiv",
    primaryClass = "astro-ph.CO",
    doi = "10.1146/annurev-nucl-111422-024107",
    journal = "Ann. Rev. Nucl. Part. Sci.",
    volume = "73",
    pages = "153--180",
    year = "2023"
}

@article{Giare:2023xoc,
    author = "Giar{\`e}, William",
    title = "{CMB Anomalies and the Hubble Tension}",
    eprint = "2305.16919",
    archivePrefix = "arXiv",
    primaryClass = "astro-ph.CO",
    month = "5",
    year = "2023"
}

@article{Hu:2023jqc,
    author = "Hu, Jian-Ping and Wang, Fa-Yin",
    title = "{Hubble Tension: The Evidence of New Physics}",
    eprint = "2302.05709",
    archivePrefix = "arXiv",
    primaryClass = "astro-ph.CO",
    doi = "10.3390/universe9020094",
    journal = "Universe",
    volume = "9",
    number = "2",
    pages = "94",
    year = "2023"
}

@article{Verde:2023lmm,
    author = {Verde, Licia and Sch{\"o}neberg, Nils and Gil-Mar{\'\i}n, H{\'e}ctor},
    title = "{A Tale of Many H0}",
    eprint = "2311.13305",
    archivePrefix = "arXiv",
    primaryClass = "astro-ph.CO",
    doi = "10.1146/annurev-astro-052622-033813",
    journal = "Ann. Rev. Astron. Astrophys.",
    volume = "62",
    number = "1",
    pages = "287--331",
    year = "2024"
}

@book{DiValentino:2024yew,
    editor = "Di Valentino, Eleonora and Brout, Dillon",
    title = "{The Hubble Constant Tension}",
    doi = "10.1007/978-981-99-0177-7",
    isbn = "978-981-99-0176-0, 978-981-99-0179-1, 978-981-99-0177-7",
    publisher = "Springer",
    series = "Springer Series in Astrophysics and Cosmology",
    year = "2024"
}

@article{Perivolaropoulos:2024yxv,
    author = "Perivolaropoulos, Leandros",
    title = "{Hubble tension or distance ladder crisis?}",
    eprint = "2408.11031",
    archivePrefix = "arXiv",
    primaryClass = "astro-ph.CO",
    doi = "10.1103/PhysRevD.110.123518",
    journal = "Phys. Rev. D",
    volume = "110",
    number = "12",
    pages = "123518",
    year = "2024"
}

@article{CosmoVerse:2025txj,
    author = "Di Valentino, Eleonora and others",
    collaboration = "CosmoVerse",
    title = "{The CosmoVerse White Paper: Addressing observational tensions in cosmology with systematics and fundamental physics}",
    eprint = "2504.01669",
    archivePrefix = "arXiv",
    primaryClass = "astro-ph.CO",
    month = "4",
    year = "2025"
}

@article{Allali:2024cji,
    author = "Allali, Itamar J. and Notari, Alessio and Rompineve, Fabrizio",
    title = "{Reduced Hubble tension in dark radiation models after DESI 2024}",
    eprint = "2404.15220",
    archivePrefix = "arXiv",
    primaryClass = "astro-ph.CO",
    doi = "10.1088/1475-7516/2025/03/023",
    journal = "JCAP",
    volume = "03",
    pages = "023",
    year = "2025"
}

@article{Giare:2024smz,
    author = "Giar{\`e}, William and Sabogal, Miguel A. and Nunes, Rafael C. and Di Valentino, Eleonora",
    title = "{Interacting Dark Energy after DESI Baryon Acoustic Oscillation Measurements}",
    eprint = "2404.15232",
    archivePrefix = "arXiv",
    primaryClass = "astro-ph.CO",
    doi = "10.1103/PhysRevLett.133.251003",
    journal = "Phys. Rev. Lett.",
    volume = "133",
    number = "25",
    pages = "251003",
    year = "2024"
}

@article{Wang:2024tjd,
    author = "Wang, Hao and Ye, Gen and Jiang, Jun-Qian and Piao, Yun-Song",
    title = "{Toward primordial gravitational waves and ns=1 in light of BICEP/Keck and DESI BAO data and the Hubble tension}",
    eprint = "2409.17879",
    archivePrefix = "arXiv",
    primaryClass = "astro-ph.CO",
    doi = "10.1103/w19x-trrq",
    journal = "Phys. Rev. D",
    volume = "111",
    number = "12",
    pages = "123505",
    year = "2025"
}

@article{Poulin:2025nfb,
    author = "Poulin, Vivian and Smith, Tristan L. and Calder{\'o}n, Rodrigo and Simon, Th{\'e}o",
    title = "{Impact of ACT DR6 and DESI DR2 for Early Dark Energy and the Hubble tension}",
    eprint = "2505.08051",
    archivePrefix = "arXiv",
    primaryClass = "astro-ph.CO",
    month = "5",
    year = "2025"
}

@article{Cuoco:2005qr,
    author = "Cuoco, Alessandro and Lesgourgues, Julien and Mangano, Gianpiero and Pastor, Sergio",
    title = "{Do observations prove that cosmological neutrinos are thermally distributed?}",
    eprint = "astro-ph/0502465",
    archivePrefix = "arXiv",
    reportNumber = "DFPD-05-A-11, DSF-02-2005, LAPTH-1088-05, IFIC-04-73",
    doi = "10.1103/PhysRevD.71.123501",
    journal = "Phys. Rev. D",
    volume = "71",
    pages = "123501",
    year = "2005"
}

@article{Farzan:2015pca,
    author = "Farzan, Yasaman and Hannestad, Steen",
    title = "{Neutrinos secretly converting to lighter particles to please both KATRIN and the cosmos}",
    eprint = "1510.02201",
    archivePrefix = "arXiv",
    primaryClass = "hep-ph",
    doi = "10.1088/1475-7516/2016/02/058",
    journal = "JCAP",
    volume = "02",
    pages = "058",
    year = "2016"
}

@article{Dvali:2016uhn,
    author = "Dvali, Gia and Funcke, Lena",
    title = "{Small neutrino masses from gravitational {\ensuremath{\theta}}-term}",
    eprint = "1602.03191",
    archivePrefix = "arXiv",
    primaryClass = "hep-ph",
    reportNumber = "MPP-2016-277, LMU-ASC-31-16",
    doi = "10.1103/PhysRevD.93.113002",
    journal = "Phys. Rev. D",
    volume = "93",
    number = "11",
    pages = "113002",
    year = "2016"
}

@article{Bellomo:2016xhl,
    author = "Bellomo, Nicola and Bellini, Emilio and Hu, Bin and Jimenez, Raul and Pena-Garay, Carlos and Verde, Licia",
    title = "{Hiding neutrino mass in modified gravity cosmologies}",
    eprint = "1612.02598",
    archivePrefix = "arXiv",
    primaryClass = "astro-ph.CO",
    doi = "10.1088/1475-7516/2017/02/043",
    journal = "JCAP",
    volume = "02",
    pages = "043",
    year = "2017"
}

@article{Lattanzi:2017ubx,
    author = "Lattanzi, Massimiliano and Gerbino, Martina",
    title = "{Status of neutrino properties and future prospects - Cosmological and astrophysical constraints}",
    eprint = "1712.07109",
    archivePrefix = "arXiv",
    primaryClass = "astro-ph.CO",
    doi = "10.3389/fphy.2017.00070",
    journal = "Front. in Phys.",
    volume = "5",
    pages = "70",
    year = "2018"
}

@article{Lorenz:2018fzb,
    author = "Lorenz, Christiane S. and Funcke, Lena and Calabrese, Erminia and Hannestad, Steen",
    title = "{Time-varying neutrino mass from a supercooled phase transition: current cosmological constraints and impact on the $\Omega_m$-$\sigma_8$ plane}",
    eprint = "1811.01991",
    archivePrefix = "arXiv",
    primaryClass = "astro-ph.CO",
    doi = "10.1103/PhysRevD.99.023501",
    journal = "Phys. Rev. D",
    volume = "99",
    number = "2",
    pages = "023501",
    year = "2019"
}

@article{Kreisch:2019yzn,
    author = "Kreisch, Christina D. and Cyr-Racine, Francis-Yan and Dor{\'e}, Olivier",
    title = "{Neutrino puzzle: Anomalies, interactions, and cosmological tensions}",
    eprint = "1902.00534",
    archivePrefix = "arXiv",
    primaryClass = "astro-ph.CO",
    doi = "10.1103/PhysRevD.101.123505",
    journal = "Phys. Rev. D",
    volume = "101",
    number = "12",
    pages = "123505",
    year = "2020"
}

@article{Oldengott:2019lke,
    author = "Oldengott, Isabel M. and Barenboim, Gabriela and Kahlen, Sarah and Salvado, Jordi and Schwarz, Dominik J.",
    title = "{How to relax the cosmological neutrino mass bound}",
    eprint = "1901.04352",
    archivePrefix = "arXiv",
    primaryClass = "astro-ph.CO",
    reportNumber = "IFIC/19-05",
    doi = "10.1088/1475-7516/2019/04/049",
    journal = "JCAP",
    volume = "04",
    pages = "049",
    year = "2019"
}

@article{Chacko:2019nej,
    author = "Chacko, Zackaria and Dev, Abhish and Du, Peizhi and Poulin, Vivian and Tsai, Yuhsin",
    title = "{Cosmological Limits on the Neutrino Mass and Lifetime}",
    eprint = "1909.05275",
    archivePrefix = "arXiv",
    primaryClass = "hep-ph",
    doi = "10.1007/JHEP04(2020)020",
    journal = "JHEP",
    volume = "04",
    pages = "020",
    year = "2020"
}

@article{Chacko:2020hmh,
    author = "Chacko, Zackaria and Dev, Abhish and Du, Peizhi and Poulin, Vivian and Tsai, Yuhsin",
    title = "{Determining the Neutrino Lifetime from Cosmology}",
    eprint = "2002.08401",
    archivePrefix = "arXiv",
    primaryClass = "astro-ph.CO",
    doi = "10.1103/PhysRevD.103.043519",
    journal = "Phys. Rev. D",
    volume = "103",
    number = "4",
    pages = "043519",
    year = "2021"
}

@article{Escudero:2020ped,
    author = "Escudero, Miguel and Lopez-Pavon, Jacobo and Rius, Nuria and Sandner, Stefan",
    title = "{Relaxing Cosmological Neutrino Mass Bounds with Unstable Neutrinos}",
    eprint = "2007.04994",
    archivePrefix = "arXiv",
    primaryClass = "hep-ph",
    reportNumber = "KCL-2020-27, FTUV-20-0625.4735, IFIC/20-33",
    doi = "10.1007/JHEP12(2020)119",
    journal = "JHEP",
    volume = "12",
    pages = "119",
    year = "2020"
}

@article{Esteban:2021ozz,
    author = "Esteban, Ivan and Salvado, Jordi",
    title = "{Long Range Interactions in Cosmology: Implications for Neutrinos}",
    eprint = "2101.05804",
    archivePrefix = "arXiv",
    primaryClass = "hep-ph",
    doi = "10.1088/1475-7516/2021/05/036",
    journal = "JCAP",
    volume = "05",
    pages = "036",
    year = "2021"
}

@article{Esteban:2022rjk,
    author = "Esteban, Ivan and Mena, Olga and Salvado, Jordi",
    title = "{Nonstandard neutrino cosmology dilutes the lensing anomaly}",
    eprint = "2202.04656",
    archivePrefix = "arXiv",
    primaryClass = "astro-ph.CO",
    doi = "10.1103/PhysRevD.106.083516",
    journal = "Phys. Rev. D",
    volume = "106",
    number = "8",
    pages = "083516",
    year = "2022"
}

@article{FrancoAbellan:2021hdb,
    author = "Franco Abell{\'a}n, Guillermo and Chacko, Zackaria and Dev, Abhish and Du, Peizhi and Poulin, Vivian and Tsai, Yuhsin",
    title = "{Improved cosmological constraints on the neutrino mass and lifetime}",
    eprint = "2112.13862",
    archivePrefix = "arXiv",
    primaryClass = "hep-ph",
    reportNumber = "FERMILAB-PUB-21-779-T",
    doi = "10.1007/JHEP08(2022)076",
    journal = "JHEP",
    volume = "08",
    pages = "076",
    year = "2022"
}

@article{Dvali:2021uvk,
    author = "Dvali, Gia and Funcke, Lena and Vachaspati, Tanmay",
    title = "{Time- and Space-Varying Neutrino Mass Matrix from Soft Topological Defects}",
    eprint = "2112.02107",
    archivePrefix = "arXiv",
    primaryClass = "hep-ph",
    reportNumber = "MIT-CTP/5355",
    doi = "10.1103/PhysRevLett.130.091601",
    journal = "Phys. Rev. Lett.",
    volume = "130",
    number = "9",
    pages = "091601",
    year = "2023"
}

@article{Alvey:2021xmq,
    author = "Alvey, James and Escudero, Miguel and Sabti, Nashwan and Schwetz, Thomas",
    title = "{Cosmic neutrino background detection in large-neutrino-mass cosmologies}",
    eprint = "2111.14870",
    archivePrefix = "arXiv",
    primaryClass = "hep-ph",
    reportNumber = "TUM-HEP-1374/21, KCL-2021-88",
    doi = "10.1103/PhysRevD.105.063501",
    journal = "Phys. Rev. D",
    volume = "105",
    number = "6",
    pages = "063501",
    year = "2022"
}

@article{Alvey:2021sji,
    author = "Alvey, James and Escudero, Miguel and Sabti, Nashwan",
    title = "{What can CMB observations tell us about the neutrino distribution function?}",
    eprint = "2111.12726",
    archivePrefix = "arXiv",
    primaryClass = "astro-ph.CO",
    reportNumber = "TUM-HEP-1375/21, KCL-2021-87",
    doi = "10.1088/1475-7516/2022/02/037",
    journal = "JCAP",
    volume = "02",
    number = "02",
    pages = "037",
    year = "2022"
}

@article{Lorenz:2021alz,
    author = {Lorenz, Christiane S. and Funcke, Lena and L{\"o}ffler, Matthias and Calabrese, Erminia},
    title = "{Reconstruction of the neutrino mass as a function of redshift}",
    eprint = "2102.13618",
    archivePrefix = "arXiv",
    primaryClass = "astro-ph.CO",
    doi = "10.1103/PhysRevD.104.123518",
    journal = "Phys. Rev. D",
    volume = "104",
    number = "12",
    pages = "123518",
    year = "2021"
}

@article{DEramo:2022nvb,
    author = "D'Eramo, Francesco and Di Valentino, Eleonora and Giar{\`e}, William and Hajkarim, Fazlollah and Melchiorri, Alessandro and Mena, Olga and Renzi, Fabrizio and Yun, Seokhoon",
    title = "{Cosmological bound on the QCD axion mass, redux}",
    eprint = "2205.07849",
    archivePrefix = "arXiv",
    primaryClass = "astro-ph.CO",
    doi = "10.1088/1475-7516/2022/09/022",
    journal = "JCAP",
    volume = "09",
    pages = "022",
    year = "2022"
}

@article{Escudero:2022gez,
    author = "Escudero, Miguel and Schwetz, Thomas and Terol-Calvo, Jorge",
    title = "{A seesaw model for large neutrino masses in concordance with cosmology}",
    eprint = "2211.01729",
    archivePrefix = "arXiv",
    primaryClass = "hep-ph",
    reportNumber = "CERN-TH-2022-180",
    doi = "10.1007/JHEP02(2023)142",
    journal = "JHEP",
    volume = "02",
    pages = "142",
    year = "2023",
    note = "[Addendum: JHEP 06, 119 (2024)]"
}

@article{Sen:2023uga,
    author = "Sen, Manibrata and Smirnov, Alexei Y.",
    title = "{Refractive neutrino masses, ultralight dark matter and cosmology}",
    eprint = "2306.15718",
    archivePrefix = "arXiv",
    primaryClass = "hep-ph",
    doi = "10.1088/1475-7516/2024/01/040",
    journal = "JCAP",
    volume = "01",
    pages = "040",
    year = "2024"
}

@article{Giare:2023qqn,
    author = "Giar{\`e}, William and G{\'o}mez-Valent, Adri{\`a} and Di Valentino, Eleonora and van de Bruck, Carsten",
    title = "{Hints of neutrino dark matter scattering in the CMB? Constraints from the marginalized and profile distributions}",
    eprint = "2311.09116",
    archivePrefix = "arXiv",
    primaryClass = "astro-ph.CO",
    doi = "10.1103/PhysRevD.109.063516",
    journal = "Phys. Rev. D",
    volume = "109",
    number = "6",
    pages = "063516",
    year = "2024"
}

@article{Brax:2023rrf,
    author = "Brax, Philippe and van de Bruck, Carsten and Di Valentino, Eleonora and Giar{\`e}, William and Trojanowski, Sebastian",
    title = "{New insights on {\ensuremath{\nu}}{\textendash}DM interactions}",
    eprint = "2303.16895",
    archivePrefix = "arXiv",
    primaryClass = "astro-ph.CO",
    doi = "10.1093/mnrasl/slad157",
    journal = "Mon. Not. Roy. Astron. Soc.",
    volume = "527",
    number = "1",
    pages = "L122--L126",
    year = "2023"
}

@article{Brax:2023tvn,
    author = "Brax, Philippe and van de Bruck, Carsten and Di Valentino, Eleonora and Giar{\`e}, William and Trojanowski, Sebastian",
    title = "{Extended analysis of neutrino-dark matter interactions with small-scale CMB experiments}",
    eprint = "2305.01383",
    archivePrefix = "arXiv",
    primaryClass = "astro-ph.CO",
    doi = "10.1016/j.dark.2023.101321",
    journal = "Phys. Dark Univ.",
    volume = "42",
    pages = "101321",
    year = "2023"
}

@article{Allali:2024anb,
    author = {Allali, Itamar J. and Aloni, Daniel and Sch{\"o}neberg, Nils},
    title = "{Cosmological probes of Dark Radiation from Neutrino Mixing}",
    eprint = "2404.16822",
    archivePrefix = "arXiv",
    primaryClass = "astro-ph.CO",
    doi = "10.1088/1475-7516/2024/09/019",
    journal = "JCAP",
    volume = "09",
    pages = "019",
    year = "2024"
}

@article{Benso:2024qrg,
    author = "Benso, Cristina and Schwetz, Thomas and Vatsyayan, Drona",
    title = "{Large neutrino mass in cosmology and keV sterile neutrino dark matter from a dark sector}",
    eprint = "2410.23926",
    archivePrefix = "arXiv",
    primaryClass = "hep-ph",
    doi = "10.1088/1475-7516/2025/04/054",
    journal = "JCAP",
    volume = "04",
    pages = "054",
    year = "2025"
}

@article{Zu:2025lrk,
    author = "Zu, Lei and Giar{\`e}, William and Zhang, Chi and Di Valentino, Eleonora and Tsai, Yue-Lin Sming and Trojanowski, Sebastian",
    title = "{Can $\nu$DM interactions solve the $S_8$ discrepancy?}",
    eprint = "2501.13785",
    archivePrefix = "arXiv",
    primaryClass = "astro-ph.CO",
    month = "1",
    year = "2025"
}

@article{Poudou:2025qcx,
    author = "Poudou, Ad{\`e}le and Simon, Th{\'e}o and Montandon, Thomas and Teixeira, Elsa M. and Poulin, Vivian",
    title = "{Self-interacting neutrinos in light of recent CMB and LSS data}",
    eprint = "2503.10485",
    archivePrefix = "arXiv",
    primaryClass = "astro-ph.CO",
    month = "3",
    year = "2025"
}

@article{Linder:2005in,
    author = "Linder, Eric V.",
    title = "{Cosmic growth history and expansion history}",
    eprint = "astro-ph/0507263",
    archivePrefix = "arXiv",
    doi = "10.1103/PhysRevD.72.043529",
    journal = "Phys. Rev. D",
    volume = "72",
    pages = "043529",
    year = "2005"
}

@article{Linder:2007hg,
    author = "Linder, Eric V. and Cahn, Robert N.",
    title = "{Parameterized Beyond-Einstein Growth}",
    eprint = "astro-ph/0701317",
    archivePrefix = "arXiv",
    doi = "10.1016/j.astropartphys.2007.09.003",
    journal = "Astropart. Phys.",
    volume = "28",
    pages = "481--488",
    year = "2007"
}

@article{Ma:1995ey,
    author = "Ma, Chung-Pei and Bertschinger, Edmund",
    title = "{Cosmological perturbation theory in the synchronous and conformal Newtonian gauges}",
    eprint = "astro-ph/9506072",
    archivePrefix = "arXiv",
    doi = "10.1086/176550",
    journal = "Astrophys. J.",
    volume = "455",
    pages = "7--25",
    year = "1995"
}

@article{Bernardeau:2001qr,
    author = "Bernardeau, F. and Colombi, S. and Gaztanaga, E. and Scoccimarro, R.",
    title = "{Large scale structure of the universe and cosmological perturbation theory}",
    eprint = "astro-ph/0112551",
    archivePrefix = "arXiv",
    reportNumber = "SACLAY-T01-142",
    doi = "10.1016/S0370-1573(02)00135-7",
    journal = "Phys. Rept.",
    volume = "367",
    pages = "1--248",
    year = "2002"
}

@article{Wang:1998gt,
    author = "Wang, Li-Min and Steinhardt, Paul J.",
    title = "{Cluster abundance constraints on quintessence models}",
    eprint = "astro-ph/9804015",
    archivePrefix = "arXiv",
    doi = "10.1086/306436",
    journal = "Astrophys. J.",
    volume = "508",
    pages = "483--490",
    year = "1998"
}

@article{Tsujikawa:2009ku,
    author = "Tsujikawa, Shinji and Gannouji, Radouane and Moraes, Bruno and Polarski, David",
    title = "{The dispersion of growth of matter perturbations in f(R) gravity}",
    eprint = "0908.2669",
    archivePrefix = "arXiv",
    primaryClass = "astro-ph.CO",
    doi = "10.1103/PhysRevD.80.084044",
    journal = "Phys. Rev. D",
    volume = "80",
    pages = "084044",
    year = "2009"
}

@article{Gannouji:2008wt,
    author = "Gannouji, R. and Moraes, B. and Polarski, D.",
    title = "{The growth of matter perturbations in f(R) models}",
    eprint = "0809.3374",
    archivePrefix = "arXiv",
    primaryClass = "astro-ph",
    doi = "10.1088/1475-7516/2009/02/034",
    journal = "JCAP",
    volume = "02",
    pages = "034",
    year = "2009"
}

@article{Kiakotou:2007pz,
    author = "Kiakotou, Angeliki and Elgaroy, Oystein and Lahav, Ofer",
    title = "{Neutrino Mass, Dark Energy, and the Linear Growth Factor}",
    eprint = "0709.0253",
    archivePrefix = "arXiv",
    primaryClass = "astro-ph",
    doi = "10.1103/PhysRevD.77.063005",
    journal = "Phys. Rev. D",
    volume = "77",
    pages = "063005",
    year = "2008"
}

@article{Lewis:2006fu,
    author = "Lewis, Antony and Challinor, Anthony",
    title = "{Weak gravitational lensing of the CMB}",
    eprint = "astro-ph/0601594",
    archivePrefix = "arXiv",
    doi = "10.1016/j.physrep.2006.03.002",
    journal = "Phys. Rept.",
    volume = "429",
    pages = "1--65",
    year = "2006"
}

@article{Kaplinghat:2003bh,
    author = "Kaplinghat, Manoj and Knox, Lloyd and Song, Yong-Seon",
    title = "{Determining neutrino mass from the CMB alone}",
    eprint = "astro-ph/0303344",
    archivePrefix = "arXiv",
    doi = "10.1103/PhysRevLett.91.241301",
    journal = "Phys. Rev. Lett.",
    volume = "91",
    pages = "241301",
    year = "2003"
}

@article{Lesgourgues:2005yv,
    author = "Lesgourgues, Julien and Perotto, Laurence and Pastor, Sergio and Piat, Michel",
    title = "{Probing neutrino masses with cmb lensing extraction}",
    eprint = "astro-ph/0511735",
    archivePrefix = "arXiv",
    reportNumber = "LAPTH-1128-05, IFIC-05-60, APC-05-90",
    doi = "10.1103/PhysRevD.73.045021",
    journal = "Phys. Rev. D",
    volume = "73",
    pages = "045021",
    year = "2006"
}

@article{Lewis:1999bs,
    author = "Lewis, Antony and Challinor, Anthony and Lasenby, Anthony",
    title = "{Efficient computation of CMB anisotropies in closed FRW models}",
    eprint = "astro-ph/9911177",
    archivePrefix = "arXiv",
    doi = "10.1086/309179",
    journal = "Astrophys. J.",
    volume = "538",
    pages = "473--476",
    year = "2000"
}

@article{Howlett:2012mh,
    author = "Howlett, Cullan and Lewis, Antony and Hall, Alex and Challinor, Anthony",
    title = "{CMB power spectrum parameter degeneracies in the era of precision cosmology}",
    eprint = "1201.3654",
    archivePrefix = "arXiv",
    primaryClass = "astro-ph.CO",
    doi = "10.1088/1475-7516/2012/04/027",
    journal = "JCAP",
    volume = "04",
    pages = "027",
    year = "2012"
}

@article{Nguyen:2023fip,
    author = "Nguyen, Nhat-Minh and Huterer, Dragan and Wen, Yuewei",
    title = "{Evidence for Suppression of Structure Growth in the Concordance Cosmological Model}",
    eprint = "2302.01331",
    archivePrefix = "arXiv",
    primaryClass = "astro-ph.CO",
    reportNumber = "LCTP-23-03",
    doi = "10.1103/PhysRevLett.131.111001",
    journal = "Phys. Rev. Lett.",
    volume = "131",
    number = "11",
    pages = "111001",
    year = "2023"
}

@article{Torrado:2020dgo,
    author = "Torrado, Jesus and Lewis, Antony",
    title = "{Cobaya: Code for Bayesian Analysis of hierarchical physical models}",
    eprint = "2005.05290",
    archivePrefix = "arXiv",
    primaryClass = "astro-ph.IM",
    reportNumber = "TTK-20-15",
    doi = "10.1088/1475-7516/2021/05/057",
    journal = "JCAP",
    volume = "05",
    pages = "057",
    year = "2021"
}

@article{Lewis:2019xzd,
    author = "Lewis, Antony",
    title = "{GetDist: a Python package for analysing Monte Carlo samples}",
    eprint = "1910.13970",
    archivePrefix = "arXiv",
    primaryClass = "astro-ph.IM",
    month = "10",
    year = "2019"
}

@article{Planck:2019nip,
    author = "Aghanim, N. and others",
    collaboration = "Planck",
    title = "{Planck 2018 results. V. CMB power spectra and likelihoods}",
    eprint = "1907.12875",
    archivePrefix = "arXiv",
    primaryClass = "astro-ph.CO",
    doi = "10.1051/0004-6361/201936386",
    journal = "Astron. Astrophys.",
    volume = "641",
    pages = "A5",
    year = "2020"
}

@article{Planck:2018yye,
    author = "Akrami, Y. and others",
    collaboration = "Planck",
    title = "{Planck 2018 results. IV. Diffuse component separation}",
    eprint = "1807.06208",
    archivePrefix = "arXiv",
    primaryClass = "astro-ph.CO",
    doi = "10.1051/0004-6361/201833881",
    journal = "Astron. Astrophys.",
    volume = "641",
    pages = "A4",
    year = "2020"
}

@article{Planck:2018lbu,
    author = "Aghanim, N. and others",
    collaboration = "Planck",
    title = "{Planck 2018 results. VIII. Gravitational lensing}",
    eprint = "1807.06210",
    archivePrefix = "arXiv",
    primaryClass = "astro-ph.CO",
    doi = "10.1051/0004-6361/201833886",
    journal = "Astron. Astrophys.",
    volume = "641",
    pages = "A8",
    year = "2020"
}

@article{Rosenberg:2022sdy,
    author = "Rosenberg, Erik and Gratton, Steven and Efstathiou, George",
    title = "{CMB power spectra and cosmological parameters from Planck PR4 with CamSpec}",
    eprint = "2205.10869",
    archivePrefix = "arXiv",
    primaryClass = "astro-ph.CO",
    doi = "10.1093/mnras/stac2744",
    journal = "Mon. Not. Roy. Astron. Soc.",
    volume = "517",
    number = "3",
    pages = "4620--4636",
    year = "2022"
}

@article{Planck:2020olo,
    author = "Akrami, Y. and others",
    collaboration = "Planck",
    title = "{$Planck$ intermediate results. LVII. Joint Planck LFI and HFI data processing}",
    eprint = "2007.04997",
    archivePrefix = "arXiv",
    primaryClass = "astro-ph.CO",
    doi = "10.1051/0004-6361/202038073",
    journal = "Astron. Astrophys.",
    volume = "643",
    pages = "A42",
    year = "2020"
}

@article{Carron:2022eyg,
    author = "Carron, Julien and Mirmelstein, Mark and Lewis, Antony",
    title = "{CMB lensing from Planck PR4~maps}",
    eprint = "2206.07773",
    archivePrefix = "arXiv",
    primaryClass = "astro-ph.CO",
    doi = "10.1088/1475-7516/2022/09/039",
    journal = "JCAP",
    volume = "09",
    pages = "039",
    year = "2022"
}

@article{Calabrese:2008rt,
    author = "Calabrese, Erminia and Slosar, Anze and Melchiorri, Alessandro and Smoot, George F. and Zahn, Oliver",
    title = "{Cosmic Microwave Weak lensing data as a test for the dark universe}",
    eprint = "0803.2309",
    archivePrefix = "arXiv",
    primaryClass = "astro-ph",
    doi = "10.1103/PhysRevD.77.123531",
    journal = "Phys. Rev. D",
    volume = "77",
    pages = "123531",
    year = "2008"
}

@article{DiValentino:2015bja,
    author = "Di Valentino, Eleonora and Melchiorri, Alessandro and Silk, Joseph",
    title = "{Cosmological hints of modified gravity?}",
    eprint = "1509.07501",
    archivePrefix = "arXiv",
    primaryClass = "astro-ph.CO",
    doi = "10.1103/PhysRevD.93.023513",
    journal = "Phys. Rev. D",
    volume = "93",
    number = "2",
    pages = "023513",
    year = "2016"
}

@article{Renzi:2017cbg,
    author = "Renzi, Fabrizio and Di Valentino, Eleonora and Melchiorri, Alessandro",
    title = "{Cornering the Planck $A_{lens}$ anomaly with future CMB data}",
    eprint = "1712.08758",
    archivePrefix = "arXiv",
    primaryClass = "astro-ph.CO",
    doi = "10.1103/PhysRevD.97.123534",
    journal = "Phys. Rev. D",
    volume = "97",
    number = "12",
    pages = "123534",
    year = "2018"
}

@article{Domenech:2020qay,
    author = "Dom{\`e}nech, Guillem and Chen, Xingang and Kamionkowski, Marc and Loeb, Abraham",
    title = "{Planck residuals anomaly as a fingerprint of alternative scenarios to inflation}",
    eprint = "2005.08998",
    archivePrefix = "arXiv",
    primaryClass = "astro-ph.CO",
    doi = "10.1088/1475-7516/2020/10/005",
    journal = "JCAP",
    volume = "10",
    pages = "005",
    year = "2020"
}

@article{DiValentino:2013mt,
    author = "Di Valentino, Eleonora and Galli, Silvia and Lattanzi, Massimiliano and Melchiorri, Alessandro and Natoli, Paolo and Pagano, Luca and Said, Najla",
    title = "{Tickling the CMB damping tail: Scrutinizing the tension between the Atacama Cosmology Telescope and South Pole Telescope experiments}",
    eprint = "1301.7343",
    archivePrefix = "arXiv",
    primaryClass = "astro-ph.CO",
    doi = "10.1103/PhysRevD.88.023501",
    journal = "Phys. Rev. D",
    volume = "88",
    number = "2",
    pages = "023501",
    year = "2013"
}

@article{Capozzi:2017ipn,
    author = "Capozzi, Francesco and Di Valentino, Eleonora and Lisi, Eligio and Marrone, Antonio and Melchiorri, Alessandro and Palazzo, Antonio",
    title = "{Global constraints on absolute neutrino masses and their ordering}",
    eprint = "2003.08511",
    archivePrefix = "arXiv",
    primaryClass = "hep-ph",
    reportNumber = "MPP-2020-30",
    doi = "10.1103/PhysRevD.95.096014",
    journal = "Phys. Rev. D",
    volume = "95",
    number = "9",
    pages = "096014",
    year = "2017",
    note = "[Addendum: Phys.Rev.D 101, 116013 (2020)]"
}

@article{DiValentino:2021imh,
    author = "Di Valentino, Eleonora and Melchiorri, Alessandro",
    title = "{Neutrino Mass Bounds in the Era of Tension Cosmology}",
    eprint = "2112.02993",
    archivePrefix = "arXiv",
    primaryClass = "astro-ph.CO",
    doi = "10.3847/2041-8213/ac6ef5",
    journal = "Astrophys. J. Lett.",
    volume = "931",
    number = "2",
    pages = "L18",
    year = "2022"
}

@article{Capozzi:2021fjo,
    author = "Capozzi, Francesco and Di Valentino, Eleonora and Lisi, Eligio and Marrone, Antonio and Melchiorri, Alessandro and Palazzo, Antonio",
    title = "{Unfinished fabric of the three neutrino paradigm}",
    eprint = "2107.00532",
    archivePrefix = "arXiv",
    primaryClass = "hep-ph",
    doi = "10.1103/PhysRevD.104.083031",
    journal = "Phys. Rev. D",
    volume = "104",
    number = "8",
    pages = "083031",
    year = "2021"
}

@article{Allali:2024aiv,
    author = "Allali, Itamar J. and Notari, Alessio",
    title = "{Neutrino mass bounds from DESI 2024 are relaxed by Planck PR4 and cosmological supernovae}",
    eprint = "2406.14554",
    archivePrefix = "arXiv",
    primaryClass = "astro-ph.CO",
    doi = "10.1088/1475-7516/2024/12/020",
    journal = "JCAP",
    volume = "12",
    pages = "020",
    year = "2024"
}

@article{Specogna:2024euz,
    author = "Specogna, Enrico and Giar{\`e}, William and Di Valentino, Eleonora",
    title = "{Planck-PR4 anisotropy spectra show better consistency with general relativity}",
    eprint = "2411.03896",
    archivePrefix = "arXiv",
    primaryClass = "astro-ph.CO",
    doi = "10.1103/PhysRevD.111.103510",
    journal = "Phys. Rev. D",
    volume = "111",
    number = "10",
    pages = "103510",
    year = "2025"
}

@article{Ishak:2024jhs,
    author = "Ishak, M. and others",
    title = "{Modified Gravity Constraints from the Full Shape Modeling of Clustering Measurements from DESI 2024}",
    eprint = "2411.12026",
    archivePrefix = "arXiv",
    primaryClass = "astro-ph.CO",
    reportNumber = "FERMILAB-PUB-24-0848-PPD",
    month = "11",
    year = "2024"
}

@article{DESI:2024hhd,
    author = "Adame, A. G. and others",
    collaboration = "DESI",
    title = "{DESI 2024 VII: Cosmological Constraints from the Full-Shape Modeling of Clustering Measurements}",
    eprint = "2411.12022",
    archivePrefix = "arXiv",
    primaryClass = "astro-ph.CO",
    reportNumber = "FERMILAB-PUB-24-0854-PPD",
    month = "11",
    year = "2024"
}

@article{DESI:2025qqy,
    author = "Andrade, U. and others",
    collaboration = "DESI",
    title = "{Validation of the DESI DR2 Measurements of Baryon Acoustic Oscillations from Galaxies and Quasars}",
    eprint = "2503.14742",
    archivePrefix = "arXiv",
    primaryClass = "astro-ph.CO",
    reportNumber = "FERMILAB-PUB-25-0162-PPD",
    month = "3",
    year = "2025"
}

@article{DESI:2025zpo,
    author = "Abdul Karim, M. and others",
    collaboration = "DESI",
    title = "{DESI DR2 Results I: Baryon Acoustic Oscillations from the Lyman Alpha Forest}",
    eprint = "2503.14739",
    archivePrefix = "arXiv",
    primaryClass = "astro-ph.CO",
    reportNumber = "FERMILAB-PUB-25-0167-PPD",
    month = "3",
    year = "2025"
}

@article{Brout:2022vxf,
    author = "Brout, Dillon and others",
    title = "{The Pantheon+ Analysis: Cosmological Constraints}",
    eprint = "2202.04077",
    archivePrefix = "arXiv",
    primaryClass = "astro-ph.CO",
    doi = "10.3847/1538-4357/ac8e04",
    journal = "Astrophys. J.",
    volume = "938",
    number = "2",
    pages = "110",
    year = "2022"
}

@article{DES:2024tys,
    author = "Abbott, T. M. C. and others",
    collaboration = "DES",
    title = "{The Dark Energy Survey: Cosmology Results with {\ensuremath{\sim}}1500 New High-redshift Type Ia Supernovae Using the Full 5 yr Data Set}",
    eprint = "2401.02929",
    archivePrefix = "arXiv",
    primaryClass = "astro-ph.CO",
    reportNumber = "FERMILAB-PUB-23-0821-PPD, DES-2023-805",
    doi = "10.3847/2041-8213/ad6f9f",
    journal = "Astrophys. J. Lett.",
    volume = "973",
    number = "1",
    pages = "L14",
    year = "2024"
}

@article{DES:2024upw,
    author = "S{\'a}nchez, B. O. and others",
    collaboration = "DES",
    title = "{The Dark Energy Survey Supernova Program: Light Curves and 5 Yr Data Release}",
    eprint = "2406.05046",
    archivePrefix = "arXiv",
    primaryClass = "astro-ph.CO",
    reportNumber = "DES-2023-0807, FERMILAB-PUB-24-0292-PPD",
    doi = "10.3847/1538-4357/ad739a",
    journal = "Astrophys. J.",
    volume = "975",
    number = "1",
    pages = "5",
    year = "2024"
}

@article{DES:2024hip,
    author = "Vincenzi, M. and others",
    collaboration = "DES",
    title = "{The Dark Energy Survey Supernova Program: Cosmological Analysis and Systematic Uncertainties}",
    eprint = "2401.02945",
    archivePrefix = "arXiv",
    primaryClass = "astro-ph.CO",
    reportNumber = "FERMILAB-PUB-23-693-PPD",
    doi = "10.3847/1538-4357/ad5e6c",
    journal = "Astrophys. J.",
    volume = "975",
    number = "1",
    pages = "86",
    year = "2024"
}

@article{Rubin:2023ovl,
    author = "Rubin, David and others",
    title = "{Union Through UNITY: Cosmology with 2,000 SNe Using a Unified Bayesian Framework}",
    eprint = "2311.12098",
    archivePrefix = "arXiv",
    primaryClass = "astro-ph.CO",
    month = "11",
    year = "2023"
}

@article{DES:2025tir,
    author = "Vincenzi, M. and others",
    collaboration = "DES",
    title = "{Comparing the DES-SN5YR and Pantheon+ SN cosmology analyses: Investigation based on ''Evolving Dark Energy or Supernovae systematics?''}",
    eprint = "2501.06664",
    archivePrefix = "arXiv",
    primaryClass = "astro-ph.CO",
    reportNumber = "FERMILAB-PUB-24-0950-PPD",
    month = "1",
    year = "2025"
}

@article{DiValentino:2020vvd,
    author = "Di Valentino, Eleonora and others",
    title = "{Cosmology Intertwined III: $f \sigma_8$ and $S_8$}",
    eprint = "2008.11285",
    archivePrefix = "arXiv",
    primaryClass = "astro-ph.CO",
    reportNumber = "FERMILAB-PUB-20-495-AE",
    doi = "10.1016/j.astropartphys.2021.102604",
    journal = "Astropart. Phys.",
    volume = "131",
    pages = "102604",
    year = "2021"
}

@article{DiValentino:2018gcu,
    author = "Di Valentino, Eleonora and Bridle, Sarah",
    title = "{Exploring the Tension between Current Cosmic Microwave Background and Cosmic Shear Data}",
    doi = "10.3390/sym10110585",
    journal = "Symmetry",
    volume = "10",
    number = "11",
    pages = "585",
    year = "2018"
}

@article{Nunes:2021ipq,
    author = "Nunes, Rafael C. and Vagnozzi, Sunny",
    title = "{Arbitrating the S8 discrepancy with growth rate measurements from redshift-space distortions}",
    eprint = "2106.01208",
    archivePrefix = "arXiv",
    primaryClass = "astro-ph.CO",
    doi = "10.1093/mnras/stab1613",
    journal = "Mon. Not. Roy. Astron. Soc.",
    volume = "505",
    number = "4",
    pages = "5427--5437",
    year = "2021"
}

@article{DES:2021bvc,
    author = "Amon, A. and others",
    collaboration = "DES",
    title = "{Dark Energy Survey Year 3 results: Cosmology from cosmic shear and robustness to data calibration}",
    eprint = "2105.13543",
    archivePrefix = "arXiv",
    primaryClass = "astro-ph.CO",
    reportNumber = "FERMILAB-PUB-21-250-AE, DES-2019-0479",
    doi = "10.1103/PhysRevD.105.023514",
    journal = "Phys. Rev. D",
    volume = "105",
    number = "2",
    pages = "023514",
    year = "2022"
}

@article{DES:2021vln,
    author = "Secco, L. F. and others",
    collaboration = "DES",
    title = "{Dark Energy Survey Year 3 results: Cosmology from cosmic shear and robustness to modeling uncertainty}",
    eprint = "2105.13544",
    archivePrefix = "arXiv",
    primaryClass = "astro-ph.CO",
    reportNumber = "FERMILAB-PUB-21-253-AE, DES-2019-0480",
    doi = "10.1103/PhysRevD.105.023515",
    journal = "Phys. Rev. D",
    volume = "105",
    number = "2",
    pages = "023515",
    year = "2022"
}

@article{KiDS:2020suj,
    author = "Asgari, Marika and others",
    collaboration = "KiDS",
    title = "{KiDS-1000 Cosmology: Cosmic shear constraints and comparison between two point statistics}",
    eprint = "2007.15633",
    archivePrefix = "arXiv",
    primaryClass = "astro-ph.CO",
    doi = "10.1051/0004-6361/202039070",
    journal = "Astron. Astrophys.",
    volume = "645",
    pages = "A104",
    year = "2021"
}

@article{Asgari:2019fkq,
    author = "Asgari, Marika and others",
    title = "{KiDS+VIKING-450 and DES-Y1 combined: Mitigating baryon feedback uncertainty with COSEBIs}",
    eprint = "1910.05336",
    archivePrefix = "arXiv",
    primaryClass = "astro-ph.CO",
    doi = "10.1051/0004-6361/201936512",
    journal = "Astron. Astrophys.",
    volume = "634",
    pages = "A127",
    year = "2020"
}

@article{Joudaki:2019pmv,
    author = "Joudaki, S. and others",
    title = "{KiDS+VIKING-450 and DES-Y1 combined: Cosmology with cosmic shear}",
    eprint = "1906.09262",
    archivePrefix = "arXiv",
    primaryClass = "astro-ph.CO",
    doi = "10.1051/0004-6361/201936154",
    journal = "Astron. Astrophys.",
    volume = "638",
    pages = "L1",
    year = "2020"
}

\appendix
\section{Full Analysis Results}
\label{Appendix-A}

\begin{table*}[htpb!]
\begin{center}
\renewcommand{\arraystretch}{1.5}
\resizebox{1.\textwidth}{!}{
\begin{tabular}{l || c c || c c }
\hline
\textbf{Parameter} 
& \multicolumn{2}{c||}{\textbf{$\Lambda$CDM+$\sum m_\nu$}} 
& \multicolumn{2}{c}{\textbf{$\Lambda$CDM+$\gamma$}} \\ 
\cline{2-5}
& \textbf{Plik+DESI+PP} & \textbf{Camspec+DESI+PP} & \textbf{Plik+DESI+PP} & \textbf{Camspec+DESI+PP} \\
\hline\hline
$ \Omega_\mathrm{b} h^2  $ & $  0.02250\pm 0.00013 $ & $  0.02229\pm 0.00012 $ & $  0.02258\pm 0.00013 $ & $  0.02237\pm 0.00012 $ \\ 
$ \Omega_\mathrm{c} h^2  $ & $  0.11824\pm 0.00067 $ & $  0.11815\pm 0.00064 $ & $  0.11734\pm 0.00066 $ & $  0.11733\pm 0.00064 $ \\ 
$ 100 \, \theta_\mathrm{MC}  $ & $  1.04116\pm 0.00028 $ & $  1.04096\pm 0.00023 $ & $  1.04121\pm 0.00028 $ & $  1.04101\pm 0.00023 $ \\ 
$ \tau  $ & $  0.0580^{+0.0066}_{-0.0075} $ & $  0.0561\pm 0.0069 $ & $  0.0498^{+0.0085}_{-0.0075} $ & $  0.0488^{+0.0085}_{-0.0073} $ \\ 
$ \log(10^{10} A_\mathrm{s})  $ & $  3.048\pm 0.014 $ & $  3.042\pm 0.014 $ & $  3.028^{+0.018}_{-0.016} $ & $  3.024^{+0.018}_{-0.015} $ \\ 
$ n_\mathrm{s}  $ & $  0.9694\pm 0.0033 $ & $  0.9676\pm 0.0034 $ & $  0.9719\pm 0.0033 $ & $  0.9698\pm 0.0035 $ \\ 
$\sum m_{\nu}$ [eV] & $ < 0.0710$ & $< 0.0643$ & $0.06$ & $0.06$ \\ 
$\gamma$ & $0.55$ & $0.55$ & $0.692\pm 0.056$ & $0.666\pm 0.052$ \\ 
\hline
$ \Omega_\mathrm{m}  $ & $  0.3009\pm 0.0037 $ & $  0.3018\pm 0.0036 $ & $  0.2990\pm 0.0038 $ & $  0.3006\pm 0.0037 $ \\ 
$ H_0  $ [Km/s/Mpc] & $  68.47\pm 0.30 $ & $  68.28\pm 0.29 $ & $  68.57\pm 0.30 $ & $  68.34\pm 0.29 $ \\ 
$S_8 $ & $  0.8167\pm 0.0090 $ & $  0.8157\pm 0.0082 $ & $  0.796\pm 0.010 $ & $  0.796\pm 0.010 $ \\ 
\hline \hline
\end{tabular} }
\end{center}
\caption{ Mean values and $68\%$~CL errors on the most relevant cosmological parameters, including the growth index $\gamma$, together with the $95\%$~CL upper limits on the total neutrino mass $\sum m_{\nu}$ arising from different combinations of cosmological data sets. }
\label{tab:results_appendix}
\end{table*}

\begin{figure*}[ht!]
   \includegraphics[width=0.95\textwidth]{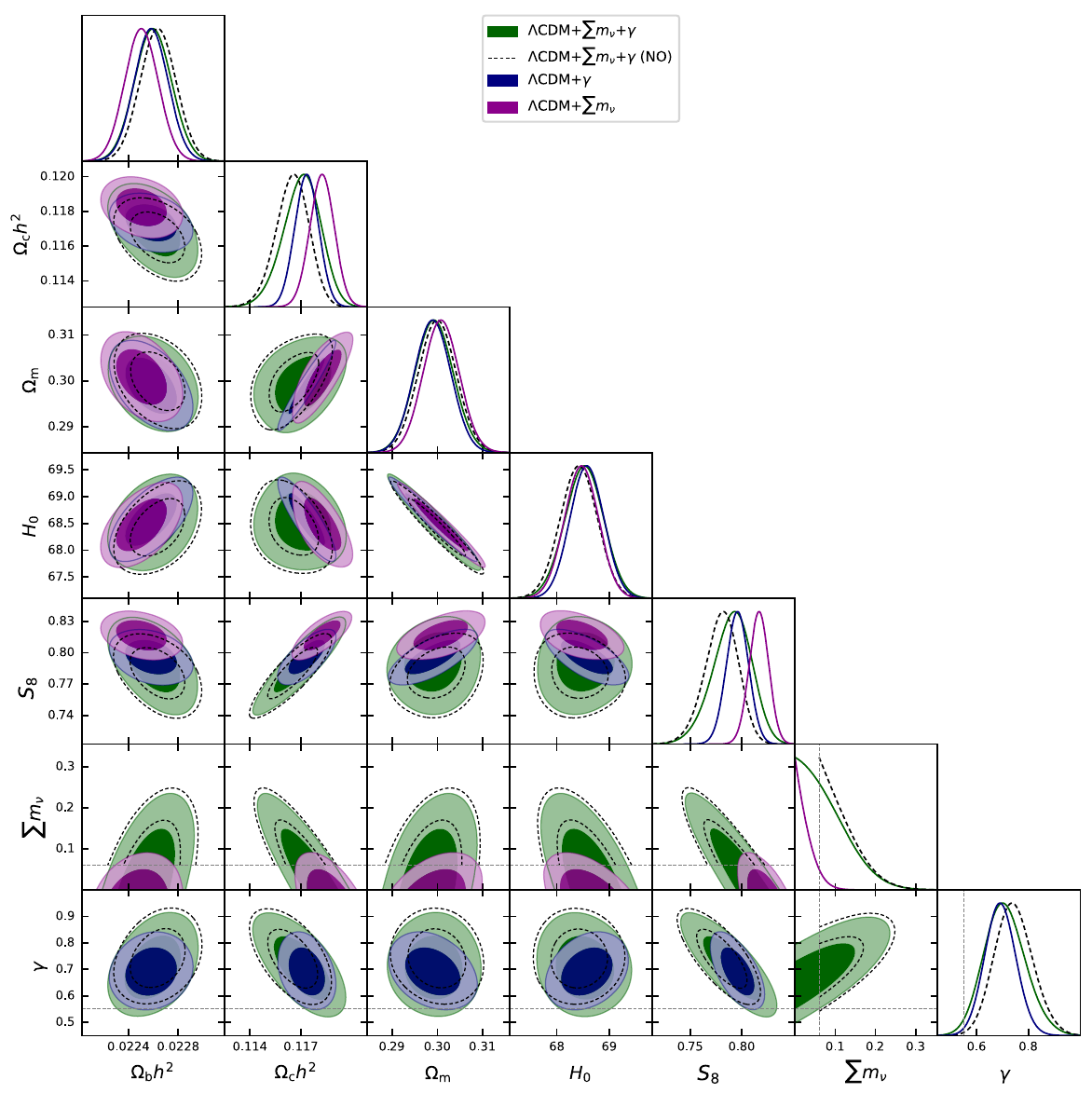}
     \caption{One-dimensional marginalized posterior distributions and two-dimensional 68\% and 95\%~CL contours for several cosmological parameters of interest, obtained from the analysis of Plik+DESI+PP under the different cosmological models and prior choices indicated in the figure legend. The dashed grey lines denote the lower bound on the total neutrino mass from oscillation experiments ($\sum m_\nu \simeq 0.06$ eV) and the standard $\Lambda$CDM prediction for the growth index ($\gamma = 0.55$).}
    \label{fig:Tplot_plik}
\end{figure*}

\begin{figure*}[ht!]
   \includegraphics[width=0.95\textwidth]{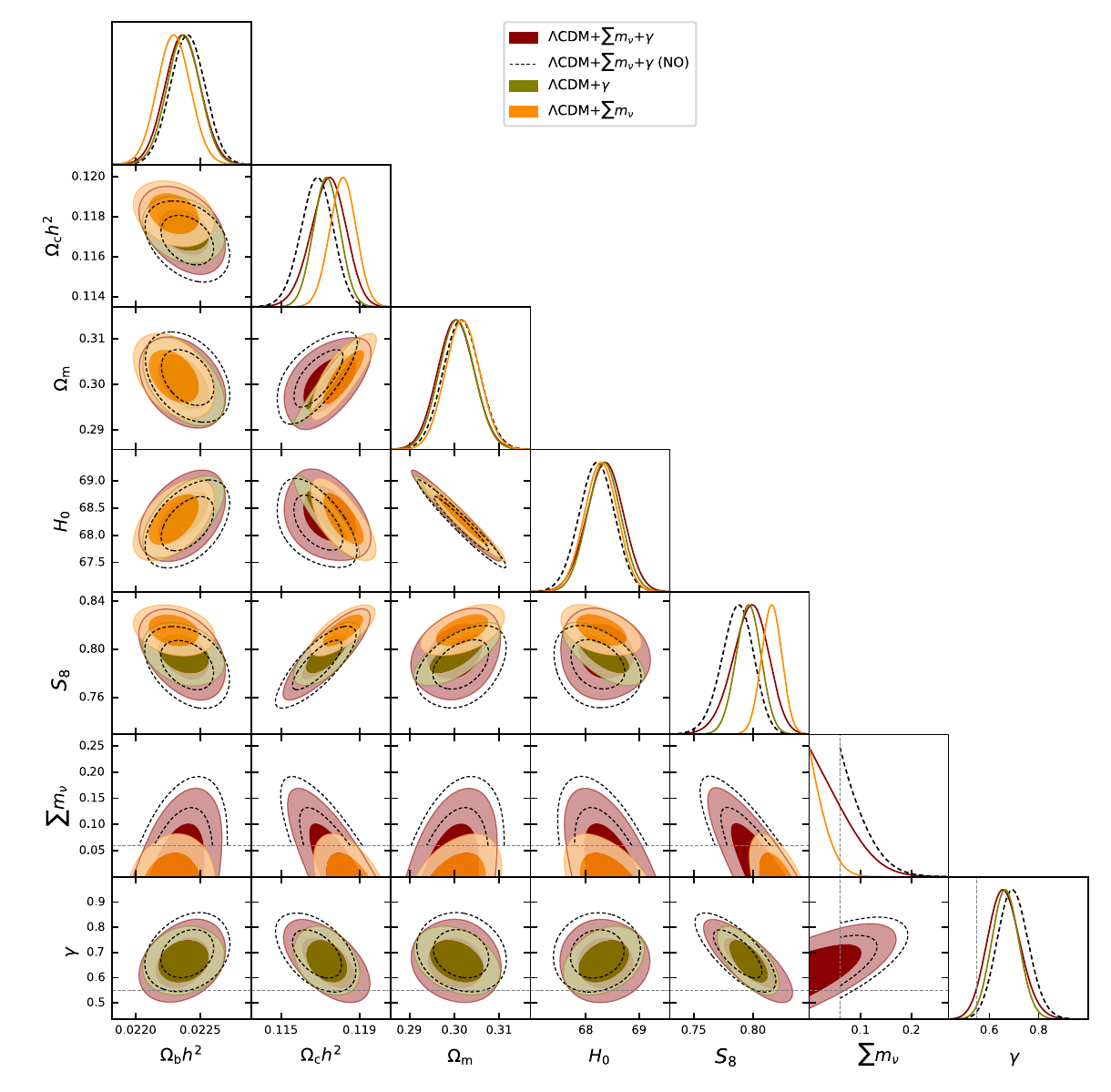}
     \caption{One-dimensional marginalized posterior distributions and two-dimensional 68\% and 95\%~CL contours for several cosmological parameters of interest, obtained from the analysis of Camspec+DESI+PP under the different cosmological models and prior choices indicated in the figure legend. The dashed grey lines denote the lower bound on the total neutrino mass from oscillation experiments ($\sum m_\nu \simeq 0.06$ eV) and the standard $\Lambda$CDM prediction for the growth index ($\gamma = 0.55$).}
    \label{fig:Tplot_camspec}
\end{figure*}

In this Appendix, we briefly illustrate the cosmological constraints on the growth index $\gamma$ and on the total neutrino mass $\sum m_\nu$ for cosmological scenarios in which only one between $\sum m_\nu$ and $\gamma$ is allowed to vary freely while the other is kept fixed in the model.

We show in Tab.~\ref{tab:results_appendix}  the main results obtained in this scenario, while Fig.~\ref{fig:Tplot_plik} and Fig.~\ref{fig:Tplot_camspec} display the one-dimensional distribution functions and the two-dimensional probability contours obtained for the different models (including the case where both $\gamma$ and $\sum m_{\nu}$ are varied simultaneously), for Plik+DESI+PP and Camspec+DESI+PP, respectively.

Focusing on the neutrino mass constraints within a universe with a standard growth index $\gamma = 0.55$, we find extremely tight bounds on $\sum m_\nu$ for both the Camspec and Plik CMB likelihoods. These 95\%~CL upper limits are $\sum m_{\nu}<0.0643$~eV for Camspec+DESI+PP, and slightly relaxed to $\sum m_{\nu}<0.0710$~eV for Plik+DESI+PP. In either case, the limits are close to the minimum bound expected from neutrino oscillation experiments, $\sum m_\nu > 0.06$~eV, hinting at a mild tension.

In the complementary case where the total neutrino mass is fixed to $\sum m_\nu = 0.06$~eV and the growth index is allowed to vary, the tendency toward larger values of $\gamma$ is confirmed. Indeed, we find $\gamma = 0.692 \pm 0.056$ for Plik+DESI+PP and $\gamma = 0.666 \pm 0.052$ for Camspec+DESI+PP, deviating from the expected value by $\sim 2.5\sigma$ and $\sim 2.2\sigma$, respectively.

As for the correlations between $\gamma$ and the remaining cosmological parameters, they behave as expected: as seen both in Fig.~\ref{fig:Tplot_plik} and Fig.~\ref{fig:Tplot_camspec} $\gamma$ is negatively correlated with $\Omega_c h^2$, due to the growth rate being parameterized as $\Omega_m(a)^\gamma$. 

\end{document}